\documentclass[fleqn,usenatbib]{mnras}

\usepackage{newtxtext,newtxmath}

\usepackage[T1]{fontenc}
\usepackage{ae,aecompl}

\usepackage{graphicx}	
\usepackage{amsmath}	
\usepackage{amssymb}	

\title[Searches for pulsar-like candidates from 3FHL unidentified objects]{Searches for Pulsar-like Candidates from Unidentified Objects in the Third Catalog of Hard {\it Fermi}-LAT (3FHL) sources with Machine Learning Techniques}

\author[C. Y. Hui et al.]{
C. Y. Hui,$^{1}$\thanks{E-mail: huichungyue@gmail.com}
Jongsu Lee,$^{2}$
K.L. Li,$^{1,3,4}$
Sangin Kim,$^{2}$
Kwangmin Oh,$^{2}$
Shengda Luo,$^{5}$
\newauthor
Alex P. Leung,$^{5}$
A. K. H. Kong,$^{4}$
J. Takata,$^{6}$
K. S. Cheng$^{7}$
\\
$^{1}$Department of Astronomy and Space Science, Chungnam National University, Daejeon 34134, Korea\\
$^{2}$Department of Space Science and Geology, Chungnam National University, Daejeon 34134, Korea\\
$^{3}$Department of Physics, UNIST, Ulsan 44919, Korea\\
$^{4}$Institute of Astronomy, National Tsing Hua University, Hsinchu, 30013, Taiwan\\
$^{5}$Faculty of Information Technology, Macau University of Science and Technology, Avenida Wai Long, Taipa, Macau\\
$^{6}$Institute of Particle physics and Astronomy, Huazhong University of Science and Technology, China\\ 
$^{7}$Department of Physics, University of Hong Kong, Pokfulam Road, Hong Kong\\
}

\date{Accepted XXX. Received YYY; in original form ZZZ}

\pubyear{2015}

\hypersetup{draft}
\begin{document}
\label{firstpage}
\pagerange{\pageref{firstpage}--\pageref{lastpage}}
\maketitle

\begin{abstract}
We report the results of searching pulsar-like candidates from the unidentified objects in the $3^{\rm rd}$ Catalog of Hard {\it Fermi}-LAT sources (3FHL).
Using a machine-learning based classification scheme with a nominal accuracy of $\sim98\%$, 
we have selected 27 pulsar-like objects from 200 unidentified 3FHL sources for an identification 
campaign. Using archival data, X-ray sources are found within the $\gamma-$ray error ellipses of 10 3FHL pulsar-like candidates.
Within the error circles of the much better constrained X-ray 
positions, we have also searched for the optical/infrared counterparts and examined their spectral energy distributions.
Among our short-listed candidates, the most 
secure identification is the association of 3FHL~J1823.3-1339 and its X-ray counterpart with the globular cluster Mercer~5. 
The $\gamma-$rays from the source can be contributed by a population of millisecond pulsars residing in the cluster. This 
makes Mercer~5 as one of the slowly growing hard $\gamma-$ray population of globular clusters with emission $>10$~GeV.  
Very recently, another candidate picked by our classification scheme, 3FHL~J1405.1-6118, has been identified 
as a new $\gamma-$ray binary with an orbital period of $13.7$~days.
Our X-ray analysis with a short {\it Chandra} observation has found a possible periodic signal candidate of $\sim1.4$~hrs and a putative 
extended X-ray tail of $\sim20$~arcsec long. Spectral energy distribution of its optical/infrared counterpart conforms
with a blackbody of $T_{\rm bb}\sim40000$~K and $R_{\rm bb}\sim12R_{\odot}$ at a distance of 7.7~kpc. This is consistent with its identification 
as an early O star as found by infrared spectroscopy.
\end{abstract}




\section{Introduction}
{\it Fermi} Gamma-ray Space Telescope has brought us into a new era of high energy astronomy 
by significantly expanding the population of $\gamma-$ray sources. 
In particular for pulsars, thanks to the much improved sensitivity of the Large Area Telescope (LAT) 
on board {\it Fermi}, our understandings 
of their high energy properties have been advanced considerably in the last decade (see \citet{hui2018} for a recent 
review). Currently, there are 234 $\gamma-$ray pulsars have been detected, 
which is $>30$ times of their population before the launch of {\it Fermi}. Not only enlarging the population, 
{\it Fermi} LAT also has uncovered previously unknown classes of $\gamma-$ray pulsars \citep{abdo2013} such as 
millisecond pulsars (MSPs). Furthermore, other $\gamma-$ray phenomena related to pulsars have 
also been found. For example, $\gamma-$ray emission were discovered from a number of globular clusters
\citep{abdo2009,kong2010,tam2011}, which 
can be originated from the collective contribution of the magnetospheric radiation from MSPs in the cluster 
\citep{abdo2010} and/or 
from the inverse Compton scattering between the relativistic pulsar wind outflow and the local soft 
photon field \citep{cheng2010,hui2011}. Also, flares in X-ray , GeV and TeV regimes from 
the $\gamma-$ray binaries, which contains a pulsar and a OB companion, were detected before/after the periastron 
passage (e.g. \citet{tam2018}).
These flares are suggested to be resulted from the intrabinary shocks \citep{takata2017}. 

In the previous {\it Fermi} LAT point source catalogs obtained from the full band all-sky survey ($>100$~MeV), 
there are approximately one-third of the sources have their nature unidentified (e.g. 2FGL; \cite{nolan2012};
3FGL: \citep{acero2015}). The locations 
of these unidentified {\it Fermi} objects provide us with a ``treasure map" for searching 
interesting objects with multiwavelength observations. By imposing a suitable set of classification criteria, 
one can select some promising candidates from these unidentified sources for searching the counterparts within their $\gamma-$ray positional 
error ellipses. For example, by choosing the unidentified objects that have low $\gamma-$ray flux variability for discriminating them from the AGN-like sources 
(i.e. small variability indices) and with curved spectral shape similar to the pulsars (i.e. large curvature significances), 
one can obtain a list of pulsar candidates for follow-up identifications \citep[e.g.][]{kong2012,hui2015,pablo2016}. 
A significant fractions of MSPs were discovered by this method \citep{clark2017}.

Apart from the full-band $\gamma-$ray source catalogs, lists of sources in the hard $\gamma-$ray bands have also been 
compiled. In the third Catalog of Hard {\it Fermi} LAT sources (3FHL) \citep{ajello2017}, it contains 1556 objects detected 
in the energy range of 10~GeV to 2~TeV. 136 of them have their nature identified and 1220 
``associated" sources have been classified primarily by the positional coincidence with sources of known nature. 
Among these 1356 sources, 59 sources are labeled as pulsars and the rest includes mostly AGNs. The remaining 200 sources 
do not have any association/identification in the 3FHL catalog.

A recent systematic investigation have been carried out for pinpointing the nature of these unidentified 3FHL objects 
\citep{kaur2019}. They have selected 110 sources from 200 unidentified 3FHL sources which have their fields  
covered by archival {\it Swift}-XRT data for their analysis. Among them, 52 sources have a single X-ray sources detected in 
their 95\% $\gamma-$ray error ellipses and have been selected for further analysis. By cross-matching 
the X-ray positions with catalogs of different wavelengths, \citet{kaur2019} have classifed 36 of these sources as 
AGN candidates. 

While their work is successful in identifying a number of AGN candidates, their approach is not very efficient as 
they have to analyze a large number of sources without any pre-screening. A lot of effort have been spent on analyzing 
the data of the sources that are unlikely to be their target-of-interest (i.e. AGN). A more efficient
approach is to select the promising candidates first with machine learning algorithms and then look into 
the archival data and/or carry follow-up observations afterward. This is the approach we adopted in our investigation. 

In this work, we present a systematic searches for pulsar-like candidates from the unassociated/unidentified 3FHL objects 
with machine learning techniques and performed a follow-up multiwavelength identification campaign. 
While the population of pulsars with energies $>100$~MeV 
has been significantly expanded, the population in the very high energy regime (VHE $>100$~GeV) remains to be rather small. 
So far only three pulsars have their pulsed emission detected at energies $>50$~GeV (cf. \citep{hui2018} for a review). 
Besides their magnetospheric radiation, interaction of the pulsar emission and/or wind particles with their surroundings can also 
produce VHE photons such as those in $\gamma$-ray binaries and globular clusters. 
The hard $\gamma-$ray pulsar-like candidates investigated in this work have the potential for enlarging VHE pulsar population and the related 
phenomena. 

\section{PSR-like Candidate Selection with Machine Learning Techniques}
Using the 3FHL sources with identified/associated nature for 
training and testing a classifier, we can perform a binary classification of 3FHL sources between 
pulsars (PSR) and non-pulsars (NON\_PSR) by employing machine learning techniques.

Among 65 features in the catalog, sixteen features were removed in the our preprocessing stage. 
Eleven features are manually removed as we believe they are not useful in determining the source nature, such as
their 3FHL names and alternative names. We have also set a threshold of removing any feature with more than 
10\% of null values, and five more features are therefore automatically removed.

We use 1356 identified/associated sources as our sample for the feature selection and building prediction models.
Among them, 1231 $\gamma-$ray sources are identified/associated with extragalactic objects, which include starburst galaxy, 
BL Lac, flat-spectrum radio quasar type of blazar, non-blazar active galaxy, narrow-line seyfert 1,
radio galaxy and blazar candidate of uncertain type. On the other hand, there are 125 $\gamma-$ray sources reside in our Galaxy. These 
Galactic 3FHL sources include pulsars, pulsar wind nebula, supernova remnant, high mass binary, binary, globular clusters and 
star formnation regions. Since we are interested in looking for the pulsar-like candidates, we perform a one-against-all 
classification. Instead of using the original labels in the catalog for identifying their nature, 
we add a column to divide them into two classes.
For all the sources identified as (or associated with) pulsars, we put them in the class of ``PSR''.
Otherwise, we label them as ``NON\_PSR''. 

In the previous work of selecting pulsar candidates from the unidentified {\it Fermi} objects, $\gamma-$ray flux variability is an important 
feature for us to distinguish the pulsar-like sources from the AGN-like sources 
\citep[e.g.][]{hui2015}. However, there is no feature for indicating 
variability in 3FHL catalog. Instead of relying on our current knowledge for differentiating the $\gamma-$ray 
properties between pulsars and the other $\gamma-$ray sources, we employ an automatic feature selection algorithm 
\citep{leung2017,2020MNRAS.492.5377L} for picking the features which can help discriminate a source is 
PSR-like or not. We achieve this by adopting 
a scheme of Recursive Feature Elimination (RFE). RFE is a backward selection method with unimportant features are 
sequentially eliminated during a recursive process \citep{leung2017}. 
With this machine-learning based technique, attributes and patterns of the data that are overlooked by human 
investigators can be highlighted.

\begin{figure}
\centering
\includegraphics[width=3.5in,height=3in]{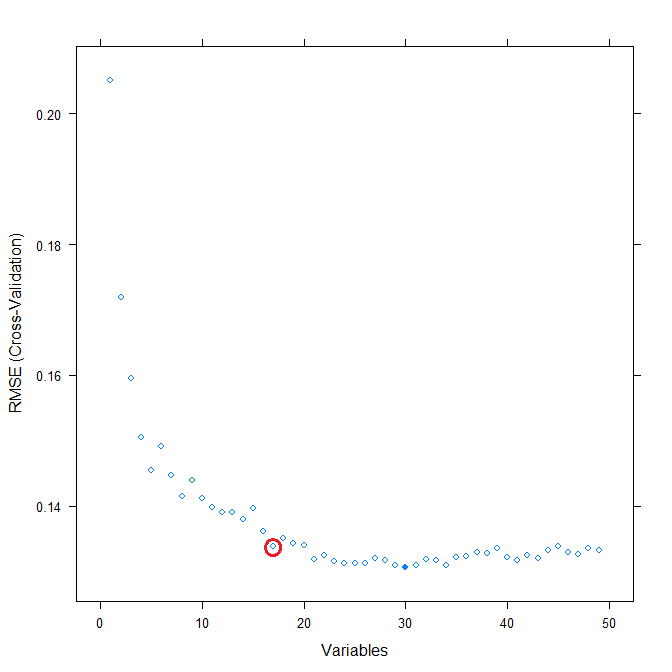}
\caption{The performance profile of PSR/NON\_PSR classification in the 3FHL catalog. The optimal performance is achieved by using thirty features (solid symbol). 
Allowing a tolerance of 1.05 as the margin of error in the RMSE value, a minimal set of 17 features are selected for building the model which is 
highlighted by the circle.}
\label{fig_profile_3fhl}
\end{figure}

After the preprocessing stage, an optimal set of features can be automatically selected by using RFE. 
For each iteration in the stage of RFE, we evaluate the performance of a random forest classifier 
by computing the root-mean-squared error (RMSE). The classification performance is evaluated by plotting 
the RMSE with the corresponding number of features. 
The performance profile produced by the RFE for the 3FHL catalog is shown in Figure~\ref{fig_profile_3fhl}.
While the minimum of the profile is attained by using 30 out of all 49 features,
it appears to be rather flat for the number of features $\gtrsim15$. The minimum corrsponds 
to 30 features (i.e. solid blue symbol in Figure~\ref{fig_profile_3fhl}) can be a result of local fluctuations. 

In view of this, we chose to suffice a little bit of performance by 
accepting an upper margin of error of $5\%$ in the RMSE value to trade for a simpler model with better interpretability.
A simpler model is often easier to understand and more robust \citep[cf.][]{2020MNRAS.492.5377L}.

With this imposed scheme, a minimal set of 17 features is selected. 
The selected features are summarized in Table~1 which are ranked by their importance scores. 

In Figure~2, we show the two-dimensional projections of the feature space for the highly ranked features:  
\texttt{Flux\_Density\_Error}, \texttt{Powerlaw\_Index} and \texttt{Pivot\_Energy}. The known PSR and NON\_PSR sources in 3FHL 
catalog are plotted as red dots and blue dots respectively. 
These chosen features suggest that the hardness of $\gamma-$rays is a key factor for differentiating PSR and 
NON\_PSR sources. This can be readily shown by the distributions the \texttt{Powerlaw\_Index}/\texttt{Spectral\_Index} (Figure~2). 
The harder a source is, the smaller these features will be. On the other hand,  
\texttt{Pivot\_Energy} is defined as the energy at which the error on differential photon flux is minimal \citep{ajello2017}. 
A softer $\gamma-$ray source has a smaller \texttt{Pivot\_Energy}, and therefore it anti-correlates with the \texttt{Powerlaw\_Index}/\texttt{Spectral\_Index}. 
For the feature \texttt{Flux\_Density\_Error}, it is the error on differential photon flux at \texttt{Pivot\_Energy} \citep{ajello2017}. 
For the hard sources, which have larger \texttt{Pivot\_Energy}, their differential fluxes 
at \texttt{Pivot\_Energy} tend to be smaller. Since \texttt{Flux\_Density\_Error} 
generally scales with the differential flux \citep[see][]{2020MNRAS.492.5377L},
this feature naturally anti-correlates with \texttt{Pivot\_Energy}.

\begin{figure*}
\centering
\includegraphics[width=4.5in]{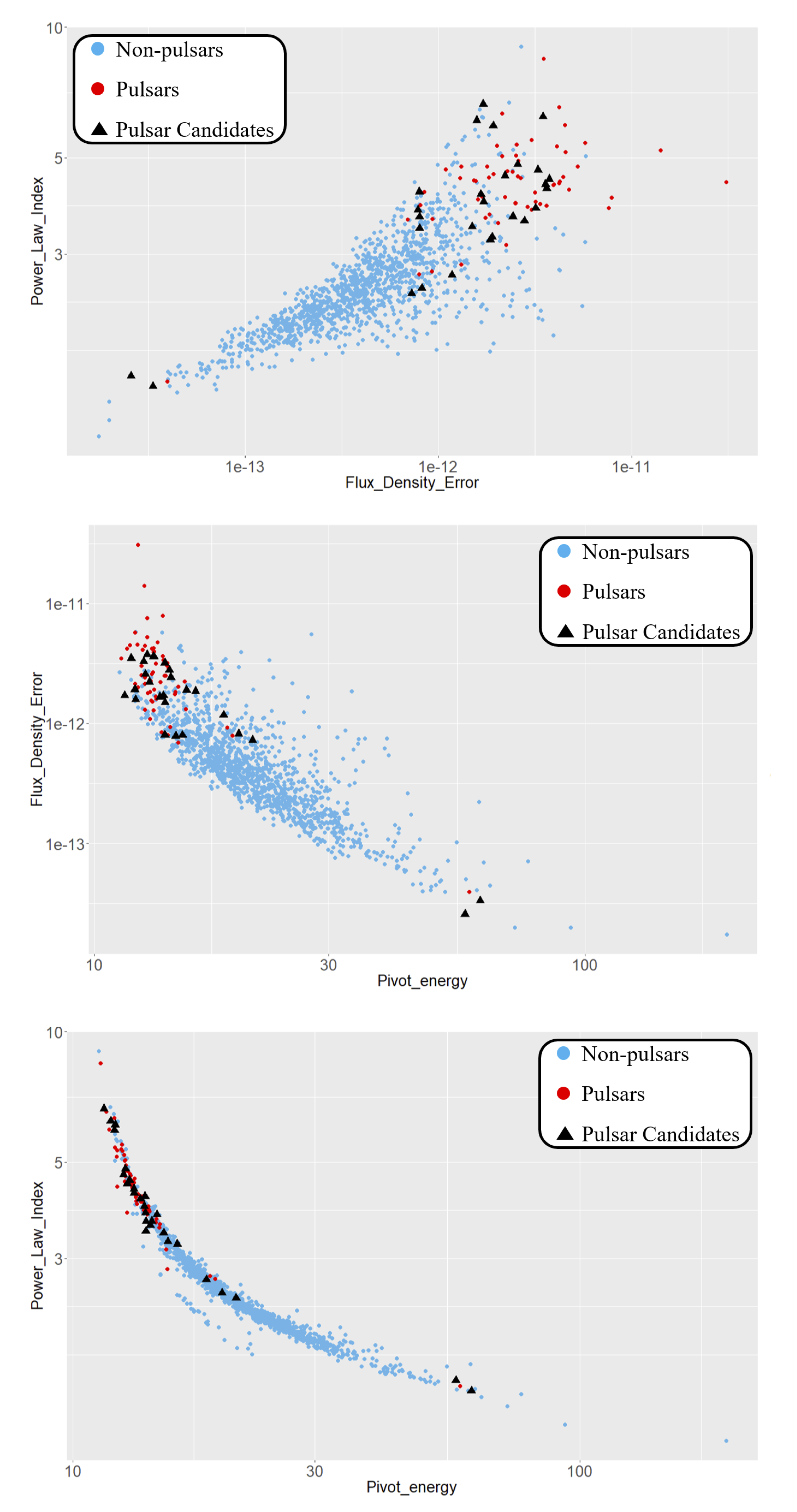}
\caption{2-dimensional projections of feature space for the selected features of high importance scores. The red dots and the blues dots show the 
distributions of known PSR sources and NON\_PSR sources in 3FHL catalog respectively. The black triangles represent the PSR candidates selected by 
our scheme.}
\end{figure*}

\begin{table}
\begin{tabular}{c | c}
\hline
\hline
Features & Importance Scores  \\
\hline
Flux\_Density\_Error       & 15.76 \\
\hline
Spectral\_Index        &  15.08 \\
\hline
Powerlaw\_Index &   14.76 \\
\hline
Pivot\_Energy   &   14.46 \\
\hline
Flux\_20\_50\_GeV\_Neg\_Err     &   8.85 \\
\hline
Powerlaw\_Index\_Error  &   7.31 \\
\hline
Flux\_20\_50\_GeV      &   7.10 \\
\hline
Flux\_Density   &   7.08     \\
\hline
BII    &   6.75 \\
\hline
Conf\_95\_SemiMinor       &   6.74   \\
\hline
Conf\_95\_SemiMajor   &   6.42 \\
\hline
Flux\_150\_500\_GeV\_Pos\_Err    &   6.35 \\
\hline
Flux\_10\_20\_GeV\_Pos\_Err     &   6.26 \\
\hline
Curve\_Significance     &   6.04 \\
\hline
Flux\_0p5\_2\_TeV       &   5.93 \\
\hline
HEP\_Prob                      &   5.93 \\
\hline
HEP\_Energy     &   5.86 \\
\hline
\end{tabular}
\label{tab_fea_rank_3fhl}
\caption{The rank of the features selected by RFE \citep{leung2017} for the 3FHL catalog. Please refer to \citet{ajello2017}
for the physical meanings of these features.}
\end{table}

One surprising result is that the feature \texttt{Curve\_Significance}, which many previous studies have relied on selecting pulsar candidates 
\citep{kong2012,hui2015,pablo2016}, 
does not appear to be a highly ranked defining characteristic for pulsars in 3FHL catalog. It has been 
found that the $\gamma-$ray spectra of most of the pulsars are characterized by a power-law with an exponential cut-off at energies 
$\lesssim5$~GeV \citep{hui2017, abdo2013}. As all the pulsars included the in 3FHL catalog are detected in the energy range of 10~GeV to 2 TeV, 
which beyonds the typical range of the spectral cut-off of most pulsars, 
their less curved spectra can be a selection effect. This may explain why \texttt{Curve\_Significance} is not among the top-ranked features 
for discriminating pulsars from the others in hard $\gamma-$ray band. 

Using the features in Table~1 to build the prediction model, we compare the performances of different classifiers.
Seven prediction models are built with the following machine learning methods:
Random Forest (RF), Generalized Additive Models (GAM),
Logistic Regression (LR), Boosted Logistic Regression (Boost LR),
Support Vector Machines (SVM),
Decision Trees (DT) and Logistic Trees (LT). For each of these tested classifiers, the data of labeled sources are 
randomly divided into training/test sets with a ratio of $70\%/30\%$. 

During the training stage, some parameters of various classifiers are tuned for optimizing their performances with the training 
data set as the input. 
Such parameters are automatically optimized by using a 10-fold cross-validation 
empirically. For quantifying 
the performance of each model, we compute the overall accuracy which is defined as the ratio of the correct 
classification in the test set.  A comparison of the overall accuracies of different classifiers is summarized 
in Table~2. Among all the tested classifiers, an optimal overall accuracy of $98.03\%$ is achieved with RF.
Using a scheme of nested cross-validation \citep{2020MNRAS.492.5377L}, we found that the standard deviations of 
all the quote accuracies in Table~2 are $\lesssim1\%$.  

\begin{table}
\centering
\begin{tabular}{c || c }
\hline
\hline
Classifiers & Our framework  \\
\hline
\textbf{RF} & \textbf{98.03$\%$}\\
\hline
Boost LR        & 97.78$\%$ \\
\hline
LR      & 97.54$\%$ \\
\hline
GAM     & 95.82$\%$ \\
\hline
SVM     & 97.54$\%$ \\
\hline
DT      & 93.31$\%$ \\
\hline
LMT     & 94.59$\%$ \\
\hline
\end{tabular}
\label{tab_accu_3fhl}
\caption{The accuracies of seven prediction models for the 3FHL catalog as evaluated by the test set.}
\end{table}

To further characterize the model performance with RF, we computed the receiver operating characteristic (ROC) curves
for the PSR/NON\_PSR classification task with both training set and test set. 
ROC curve is a plot of sensitivity (i.e. probability of detection) against specificity (i.e. 1-probability of false alarm). 
A good model should minimize the false alarm and avoid missing any detection, and hence its ROC curve would
be pushed toward the top-left corner. The training and test ROC curves of RF classifier are shown in Figure~3. 
The Area Under the Curve (AUC) of an ROC curve provides another measure for the classification performance 
The larger the AUC, the better the performance. An AUC of $98.2\%$ is obtained for the test ROC. 

\begin{figure}
\centering
\begin{minipage}{3in}
\includegraphics[width=2.5in]{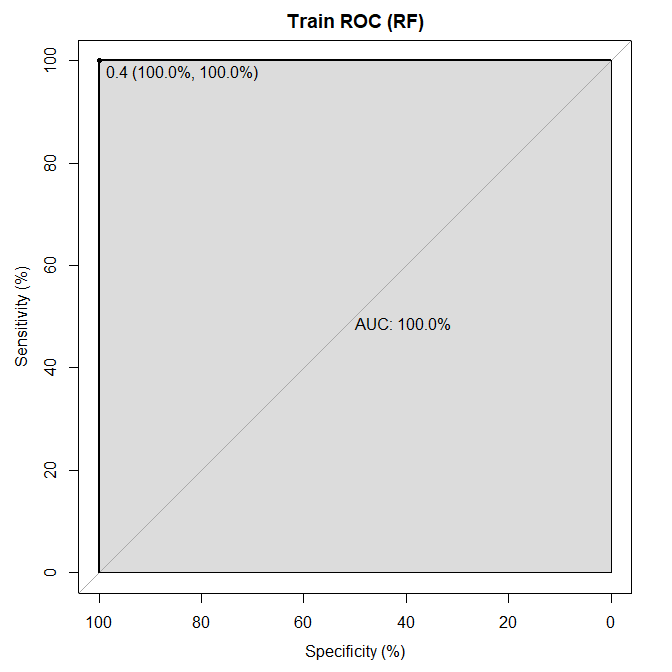}
\end{minipage}

~~~~
\begin{minipage}{3in}
\includegraphics[width=2.5in]{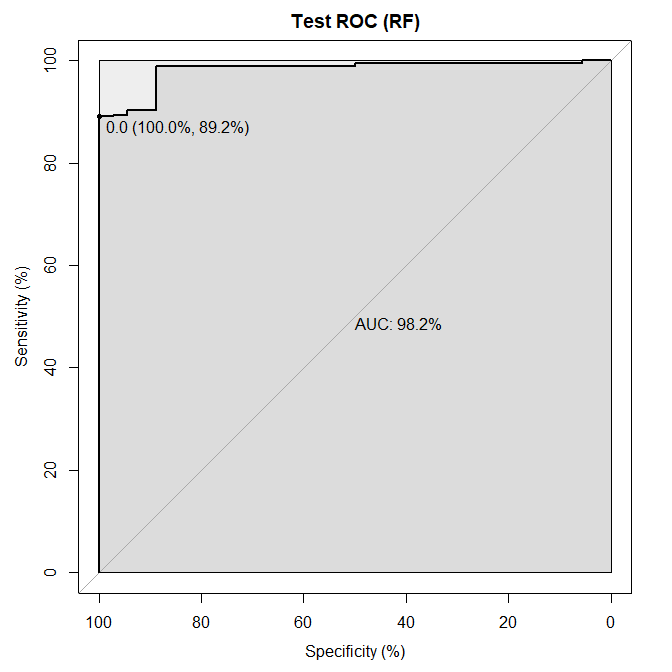}
\end{minipage}
\caption{The training and test ROC curves produced by random forest classifier using 3FHL catalog}
\label{fig_ROC_3fhl_rf}
\end{figure}

Using the prediction model with the best threshold obtained from the test ROC curve in Figure~3, we run the
PSR/NON\_PSR classification on the 200 unidentified/unassociated sources. 27 of them have been classified as PSR by our model.
We summarize their properties in Table~3, which includes their names in 3FHL catalog, $\gamma-$ray positions and errors,
the corresponding name in 3FGL catalog (if there is any), as well as the confidence score of belonging to PSR class 
assigned by our model. The confidence score for a given source provides a gauge for the reliability of the class assignment as 
predicted by the model, which should not be interpreted as the probability of the source as a PSR. 
The distributions of these 27 selected PSR candidates in the projected feature spaces are shown by the black triangles in Figure~2. 
Except for two outliers (3FHL~J1915.2-1323 and 3FHL~J0737.5+6534) with low PSR confidence scores, other candidates are clustered in 
the regime occupied by the known pulsars in 3FHL catalog.

\begin{table*}
\centering
\begin{tabular}{lccccc}
\hline
\hline
3FHL name & RA (J2000) & Dec (J2000) & $\theta_{95}$ & 3FGL name & PSR confidence score \\
\hline
          & h m s  & d m s  & degree & &\\
\hline
3FHL J1748.6-2816 &  17 48 38.4 & -28 16 41 & 0.035 &  3FGL J1748.3-2815c   &  0.858 \\
3FHL J1839.4-0553 &  18 39 24.3 & -05 53 46 & 0.040 &  3FGL J1839.3-0552    &   0.852 \\
3FHL J1823.3-1339 &  18 23 21.7 & -13 39 46 & 0.031 &  3FGL J1823.2-1339    &   0.84 \\
3FHL J1748.1-2903 &  17 48 08.9 & -29 03 42 & 0.039 &  3FGL J1747.7-2904    &  0.79 \\
3FHL J1139.2-6248 &  11 39 16.7 & -62 48 09 & 0.044 &  3FGL J1139.0-6244    &   0.768 \\
3FHL J1857.0+0059 &  18 57 05.7 & +00 59 23 & 0.058 &  3FGL J1857.2+0059    &   0.668 \\
3FHL J1753.8-2537 &  17 53 48.1 & -25 37 46 & 0.026 &  3FGL J1754.0-2538    &   0.66 \\
3FHL J1802.3-3043 &  18 02 23.7 & -30 43 20 & 0.049 &  3FGL J1802.4-3043    &   0.658   \\
3FHL J1907.0+0713 &  19 07 00.6 & +07 13 43 & 0.046 &        ...            &   0.602 \\
3FHL J1800.7-2357 &  18 00 44.1 & -23 57 12 & 0.051 &  3FGL J1800.8-2402    &   0.59 \\
3FHL J1603.3-6011 &  16 03 22.9 & -60 11 59 & 0.048 &  3FGL J1603.7-6011    &   0.522 \\
3FHL J1855.5+0142 &  18 55 35.8 & +01 42 55 & 0.047 &            ...        &   0.516 \\
3FHL J1405.1-6118 &  14 05 06.2 & -61 18 06 & 0.034 &  3FGL J1405.1-6119    &   0.466 \\
3FHL J1306.3-6042 &  13 06 22.5 & -60 42 39 & 0.028 &  3FGL J1306.4-6043    &  0.45\\
3FHL J1626.3-4915 &  16 26 23.9 & -49 15 23 & 0.076 &  3FGL J1626.2-4911    &   0.418 \\
3FHL J1112.5-6054 &  11 12 33.5 & -60 54 40 & 0.072 &  3FGL J1111.9-6058    &   0.4 \\
3FHL J1747.2-2822 &  17 47 17.7 & -28 22 00 & 0.033 &           ...        &   0.372 \\
3FHL J1824.3-0621 &  18 24 18.0 & -06 21 05 & 0.045 &  3FGL J1824.3-0620    &   0.364 \\
3FHL J0725.6-5008 &  07 25 39.1 & -50 08 25 & 0.043 &  3FGL J0725.4-5007    &   0.334 \\
3FHL J0541.1-4855 &  05 41 10.7 & -48 55 43 & 0.072 &             ...       &   0.166 \\
3FHL J1915.2-1323 &  19 15 16.4 & -13 23 30 & 0.051 &            ...        &  0.162 \\
3FHL J0737.5+6534 &  07 37 35.3 & +65 34 43 & 0.033 &           ...         &   0.158 \\
3FHL J1657.6-4656 &  16 57 37.1 & -46 56 54 & 0.095 &  3FGL J1657.6-4653    &  0.144\\
3FHL J1803.1-6709 &  18 03 10.7 & -67 09 49 & 0.053 &  3FGL J1803.3-6706    &  0.142 \\
3FHL J1200.3+0201 &  12 00 22.7 & +02 01 44 & 0.061 &  3FGL J1200.4+0202    &   0.132\\
3FHL J0110.9+4346 &  01 10 56.5 & +43 46 54 & 0.082 &        ...           &   0.124\\
3FHL J0115.4-2916 &  01 15 24.2 & -29 16 57 & 0.055 &       ...             &   0.118 \\
\hline
\end{tabular}
\label{tab_fhl_psr_candid}
\caption{27 PSR candidates selected from 3FHL catalog. $\theta_{95}$ are their 
$\gamma-$ray positional uncertainty at 95\% confidence level.}
\end{table*}

\begin{table*}
\begin{center}
\caption{Properties of X-ray Sources within $\gamma$-ray Error Ellipses (95\% Confidence) of Selected 3FHL Unidentified Objects}
\footnotesize
\begin{tabular}{c c c c c c c c c c}
\hline
\hline
Source & R.A. (J2000) & Decl. (J2000) & $\sigma_{\rm pos}$ & Signifi. & Inst. & Counts rate & F$^{\rm unabs}_{0.3-10 keV}$ & F$_X$/F$_\gamma$ & Variability \\
 & (h m s) & (d m s) & (arcsec) & ($\sigma$) &  & (10$^{-3}$ cts/s) & (10$^{-14}$ erg cm$^{-2}$ s$^{-1}$) & (10$^{-3}$) & (S / L)\\
\hline
\multicolumn{10}{c}{\bf 3FHL J1748.6-2816} \\
\hline
J17486\_X1 & 17:48:40.85 & -28:18:22.80 & 0.47 & 6.86 & C$^*$,X & 1.45$\pm0.36$ & 6.09$\pm1.51$ & $  15.20 ^{+ 9.32 }_{- 5.88 } $ & N / $2.4\sigma$ \\
\hline
\multicolumn{10}{c}{\bf 3FHL J1839.4-0553} \\
\hline
J18394\_X1 & 18:39:19.12 & -05:54:06.73 & 0.34 & 9.68 & C$^*$ & 1.56$\pm0.30$ & $ 7.69 ^{+ 1.48 }_{- 1.53 }$ & $ 20.17^{+ 11.99 }_{- 7.26 } $ & N / $1.4\sigma$ \\
J18394\_X2 & 18:39:21.02 & -05:53:28.23 & 0.41 & 7.93 & C$^*$ & 1.35$\pm0.29$ & $ 6.65 ^{+ 1.43 }_{- 1.38 }$ & $ 17.46^{+ 10.9 }_{- 6.41 } $ & N / $\cdot$ \\
\hline
\multicolumn{10}{c}{\bf 3FHL J1823.3-1339} \\
\hline
J18233\_X1 & 18:23:19.80 & -13:40:09.90 & 0.58 & 7.43 & X$^*$ & 5.06$\pm0.84$ & $ 19.82 ^{+ 3.25 }_{- 3.29 }$ & $ 34.23^{+ 15.49 }_{- 10.41 } $ & N / $\cdot$ \\
\hline
\multicolumn{10}{c}{\bf 3FHL J1748.1-2903} \\
\hline
J17481\_X1 & 17:48:04.36 & -29:02:27.97 & 0.30 & 4.02 & C$^*$ & 0.28$\pm0.09$ & $ 1.14 ^{+ 0.36 }_{- 0.37 }$ & $ 5.22 ^{+ 4.65 }_{- 2.50 }$ & N / $>2.9\sigma$ \\
J17481\_X2 & 17:48:05.67 & -29:04:05.09 & 0.16 & 8.05 & C$^*$ & 1.33$\pm0.35$ & $ 5.40 \pm1.42$ & $ 24.77 ^{+ 20.02 }_{- 10.75 }$ & N / $>2.0\sigma$ \\
\hline
\multicolumn{10}{c}{\bf 3FHL J1857.0+0059} \\
\hline
J18570\_X1 & 18:57:13.48 & +01:01:46.54 & 0.45 & 6.63 & X$^*$ & 0.68$\pm0.19$ & $2.09\pm0.57 $ & $ 6.65^{+ 4.92 }_{- 2.84 } $ & N / $\cdot$ \\
\hline
\multicolumn{10}{c}{\bf 3FHL J1800.7-2357} \\
\hline
J18007\_X1 & 18:00:36.99 & -23:55:55.90 & 0.22 & 6.46 & C$^*$ & 0.34$\pm0.08$ & 1.36$\pm0.30$ & $ 3.51^{+ 2.71 }_{- 1.43} $ & Y / $\cdot$ \\
J18007\_X2 & 18:00:33.80 & -23:57:38.24 & 0.37 & 5.91 & C$^*$ & 0.25$\pm0.06$ & 1.00$\pm0.25$ & $ 2.58^{+ 2.09 }_{- 1.10 }$ & N / $\cdot$ \\
J18007\_X3 & 18:00:49.51 & -23:59:41.61 & 0.34 & 5.42 & C$^*$ & 0.27$\pm0.07$ & 1.09$\pm0.27$ & $ 2.80 ^{+ 2.27 }_{- 1.20 }$ & N / $\cdot$ \\
J18007\_X4 & 18:00:41.89 & -23:56:48.43 & 0.52 & 5.24 & C$^*$ & 0.28$\pm0.07$ & 1.11$\pm0.28$ & $ 2.86 ^{+ 2.34 }_{- 1.23 }$ & N / $\cdot$ \\
J18007\_X5 & 18:00:32.56 & -23:58:22.45 & 0.34 & 5.09 & C$^*$ & 0.23$\pm0.06$ & 0.93$\pm0.25$ & $ 2.40 ^{+ 2.01 }_{- 1.06 }$ & N / $\cdot$ \\
J18007\_X6 & 18:00:36.88 & -23:59:24.06 & 0.22 & 5.01 & C$^*$ & 0.21$\pm0.06$ & 0.82$\pm0.23$ & $ 2.12 ^{+ 1.80 }_{- 0.95 }$ & N / $\cdot$ \\
J18007\_X7 & 18:00:44.13 & -23:57:51.04 & 0.31 & 4.35 & C$^*$ & 0.20$\pm0.06$ & 0.79$\pm0.24$ & $ 2.04 ^{+ 1.79 }_{- 0.95 }$ & N / $\cdot$ \\
J18007\_X8 & 18:00:35.67 & -23:57:50.23 & 0.34 & 4.13 & C$^*$ & 0.16$\pm0.05$ & 0.65$\pm0.21$ & $ 1.68 ^{+ 1.53 }_{- 0.81 }$ & N / $\cdot$ \\
\hline
\multicolumn{10}{c}{\bf 3FHL J1405.1-6118} \\
\hline
J14051\_X1 & 14:05:14.45 & -61:18:27.63 & 0.10 & 23.92 & C$^*$ & 5.32$\pm0.65$ & 27.09$\pm{3.31}$ & $ 48.29 ^{+ 42.19 }_{- 18.03 }$ & N / $\cdot$ \\
J14051\_X2 & 14:05:06.47 & -61:16:23.56 & 0.38 & 7.47 & C$^*$ & 1.48$\pm0.35$ & 7.54$\pm1.80$ & $ 13.43 ^{+ 14.30 }_{- 6.11 }$ & N / $\cdot$ \\
\hline
\multicolumn{10}{c}{\bf 3FHL J1626.3-4915} \\
\hline
J16263\_X1 & 16:26:01.95 & -49:14:11.29 & 0.35 & 11.77 & C$^*$,X & 6.63$\pm0.93$ & $ 33.76 ^{+ 4.73 }_{- 4.74 }$ & $ 66.46 ^{+ 65.36 }_{- 26.37 }$ & N / $2.0\sigma$ \\
J16263\_X2 & 16:26:29.28 & -49:15:43.69 & 0.33 & 8.94 & C$^*$ & 2.32$\pm0.50$ & $ 12.12\pm{ 2.61 }$ &$ 23.86 ^{+ 26.59 }_{- 10.72 }$ & N / $>3.7\sigma$ \\
J16263\_X3 & 16:26:08.53 & -49:17:44.38 & 0.55 & 6.36 & X$^*$ & 4.64$\pm1.06$ & $ 20.44 ^{+ 4.66 }_{- 4.63 }$ & $ 40.24 ^{+ 45.72 }_{- 18.40 }$ & N / $\cdot$ \\
\hline
\multicolumn{10}{c}{\bf 3FHL J1747.2-2822} \\
\hline
J17472\_X1 & 17:47:20.91 & -28:23:04.49 & 0.36 & 9.35 & C$^*$,X & 0.90$\pm0.13$ & $ 3.73 ^{+ 0.52 }_{- 0.53 }$ & $ 6.65 ^{+ 4.96 }_{- 2.42 }$ & N / $3.1\sigma$ \\
J17472\_X2 & 17:47:22.41 & -28:23:26.38 & 0.33 & 6.47 & C$^*$ & 0.47$\pm0.09$ & 1.93$\pm0.38$ & $ 3.44 ^{+ 2.87 }_{- 1.39 }$ & N / $3.3\sigma$ \\
J17472\_X3 & 17:47:09.34 & -28:21:37.01 & 0.32 & 6.18 & C$^*$ & 0.47$\pm0.09$ & 1.87$\pm0.38$ & $ 3.34 ^{+ 2.81 }_{- 1.36 }$ & N / $>2.6\sigma$ \\
J17472\_X4 & 17:47:09.33 & -28:21:55.24 & 0.26 & 5.76 & C$^*$ & 0.43$\pm0.09$ & 1.74$\pm0.37$ &$ 3.10 ^{+ 2.66 }_{- 1.29 }$ & N / $>2.4\sigma$ \\
J17472\_X5 & 17:47:14.27 & -28:21:09.82 & 0.51 & 5.11 & C$^*$ & 0.34$\pm0.08$ & 1.62$\pm0.33$ & $ 2.89 ^{+ 2.44 }_{- 1.18 }$ & N / $4.3\sigma$ \\
J17472\_X6 & 17:47:13.07 & -28:23:22.76 & 0.23 & 4.30 & C$^*$ & 0.30$\pm0.08$ & 1.22$\pm{0.32}$ & $ $ $ 2.18 ^{+ 2.05 }_{- 0.99 }$ & N / $2.5\sigma$ \\
J17472\_X7 & 17:47:20.50 & -28:23:46.16 & 0.36 & 4.28 & C$^*$ & 0.37$\pm0.10$ & 1.53$\pm0.40$ & $ 2.72 ^{+ 2.55 }_{- 1.23 }$ & N / $>1.8\sigma$ \\
J17472\_X8 & 17:47:23.80 & -28:22:30.27 & 0.45 & 4.02 & C$^*$ & 0.21$\pm0.06$ & 0.88$\pm0.26$ & $ 1.57 ^{+ 1.54 }_{- 0.74 }$ & N / $>1.5\sigma$ \\
\hline
\multicolumn{10}{c}{\bf 3FHL J0737.5+6534} \\
\hline
J07375\_X1 & 07:37:33.39 & +65:33:07.54 & 0.07 & 71.24 & C$^*$,X & 6.30$\pm0.36$ & $ 10.16\pm0.58 $ & $ 55.22 ^{+ 147.42 }_{- 24.80 }$ & N / $12.2\sigma$ \\
J07375\_X2 & 07:37:40.49 & +65:35:21.72 & 0.08 & 33.45 & C$^*$,X & 2.30$\pm0.22$ & $ 3.72 ^{+ 0.35 }_{- 0.34 }$ & $ 20.2 ^{+ 56.62 }_{- 9.48 }$ & N / $8.1\sigma$ \\
J07375\_X3 & 07:37:52.85 & +65:34:07.68 & 0.23 & 10.45 & C$^*$ & 0.66$\pm0.12$ & $ 1.06 ^{+ 0.20 }_{- 0.19 }$ & $ 5.78 ^{+ 17.96 }_{- 3.02 }$ & N / $>3.8\sigma$ \\
J07375\_X4 & 07:37:36.50 & +65:32:51.04 & 0.17 & 5.55 & C$^*$ & 0.37$\pm0.10$ & $ 0.60 \pm 0.15 $ & $ 3.25 ^{+ 10.95 }_{- 1.84 }$ & N / $>2.1\sigma$ \\
J07375\_X5 & 07:37:38.37 & +65:36:29.32 & 0.33 & 5.03 & C$^*$,X & 0.27$\pm0.08$ & $ 0.44\pm0.13 $ & $ 2.36 ^{+ 8.22 }_{- 1.38 }$ & N / $4.0\sigma$ \\
J07375\_X6 & 07:37:17.94 & +65:35:09.74 & 0.33 & 4.48 & C$^*$ & 0.27$\pm0.08$ & $ 0.43\pm0.13 $ & $ 2.36 ^{+ 8.31 }_{- 1.40 }$ & N / $>2.3\sigma$ \\
J07375\_X7 & 07:37:35.25 & +65:35:50.00 & 0.15 & 4.20 & C$^*$ & 0.25$\pm0.08$ & $ 0.40\pm0.13 $ & $ 2.18 ^{+ 7.78 }_{- 1.31 }$ & N / $>2.3\sigma$ \\
J07375\_X8 & 07:37:43.71 & +65:33:45.70 & 0.20 & 4.12 & C$^*$ & 0.28$\pm0.09$ & $ 0.46\pm0.14 $ & $ 2.48 ^{+ 8.79 }_{- 1.48 }$ & N / $>2.3\sigma$ \\
\hline
\end{tabular}
\end{center}
Note. The chance coincidence can be seen next to the 3FHL name. 
* means that the coordinate and significance are extracted from. 
For whole analysis, we used the $\texttt{wavdetect}$ task in $\texttt{CIAO}$. 
The counts rate was calculated from the net\_counts for *'s instrument with exposure time.
\end{table*}

\begin{figure*}
\centering
\includegraphics[width=9.3cm]{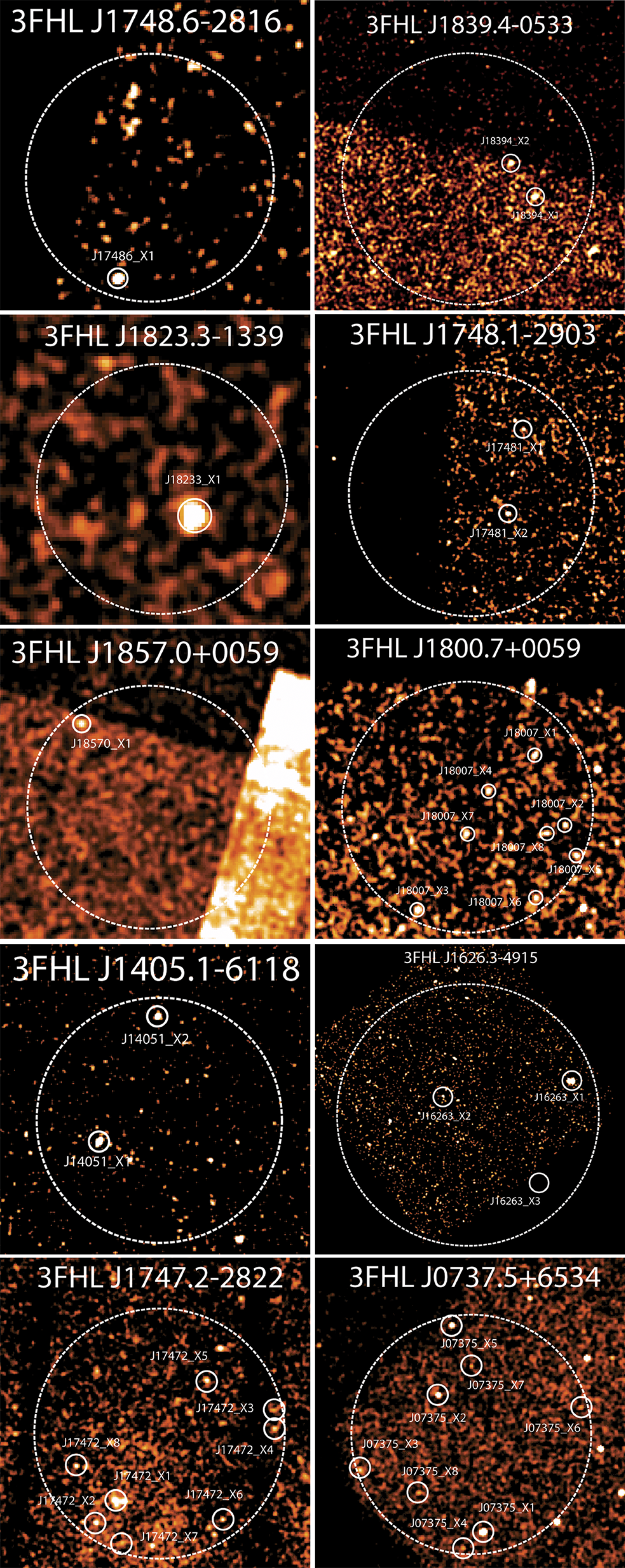}
\caption{X-ray sources (solid circles) found within the 95\% $\gamma-$ray positional error ellipses 
(dashed ellipses)of our selected PSR-like 3FHL sources. Top is north and left is east in all images.}
\end{figure*}

\citet{kaur2019} have reported 36 unidentified 3FHL sources which most likely belong to AGNs family.
In comparing their list (Table~4 in their paper) with our PSR candidate list, only one source 3FHL~J0541.1-4855, which has a relatively low PSR confidence 
score of 0.166, is overlapped. This provides further confidence for our method and the PSR candidates selected by this scheme.
 
\section{Data Analysis}
\subsection{Searching for X-ray/optical sources within the $\gamma-$ray error ellipses}
We have searched for X-ray counterparts associated with our short-listed 3FHL sources by using archival X-ray spectral
imaging data. We attempted to detect the X-ray sources within the $\gamma-$ray error ellipses of these 
candidates with a wavelet detection algorithm. 
Only the X-ray sources detected at a significance larger than $4\sigma$ are considered as genuine in our work. 
Among 27 PSR-like candidates in Table~3, ten of them have X-ray sources found within their 95\% confidence 
$\gamma$-ray error ellipses.
The results are summarized in Table~4. X-ray images of the fields of these 10 PSR-like candidates are shown in Figure~4. 

We found that these selected candidates have been observed either by {\it Chandra}, {\it XMM-Newton} or {\it Swift}. 
If {\it Chandra} data is available for a PSR-like candidate, we solely used its data to determine the positions of the X-ray counterparts 
as {\it Chandra} can provide the best positional accuracy among all X-ray telescopes. 
For the cases there is no archival {\it Chandra} data but with {\it XMM-Newton} available, the positions of the X-ray counterparts 
are determined by the MOS cameras (merged MOS1/2 data) because their pixel size provide a full sampling of the point spread function of the mirror.
For {\it Swift} XRT observations, we noticed that their exposures are typically a few ks which are unconstraining for our searches of 
relatively faint sources potentially associated with pulsars. Therefore, the observations 
by {\it Swift} XRT will be ignored in this work. 

By assuming an absorbed power-law with a photon index of $\Gamma_{x}=2$ and the column absorption $n_{H}$ adopted at the value of the 
Galactic HI column density in the directions towards these X-ray sources \citep{kalberla2005}, with the aid of PIMMS (ver. 4.9a), 
we systematically computed the absorption-corrected X-ray fluxes $F_{x}$ for all the X-ray sources in an energy
range of $0.3-10$~keV by using their count rates. And hence, we obtained their X-ray to $\gamma-$ray flux ratios $F_{x}/F_{\gamma}$ with 
$F_{\gamma}$ as the energy flux in 10~GeV to 2 TeV as obtained from 3FHL catalog. $F_{x}$ and $F_{x}/F_{\gamma}$ are summarized in 
column 8 and column 9 in Table~4.

In Figure~5, we compare the distributions of $\log F_{x}/F_{\gamma}$ of these X-ray sources with those of the known pulsars in the same 
energy ranges. The range of $\log F_{x}/F_{\gamma}$ spanned by these X-ray sources is bracketed by those of the known pulsars, except for 
two sources J18007\_X8 and J17472\_X8 which have the lowest $\log F_{x}/F_{\gamma}$. 

We have also examined the temporal variability of these X-ray sources. 
We first search for the short-term variability within each observation window
by using the Gregory-Loredo variability algorithm \citep{gl1992}. 
By testing whether the arrival times of these sources are uniformly distributed, 
only J18007\_X1 has a probability of $>90\%$ as a variable source. 

Apart from the short-term variabilities, a number of X-ray sources have been observed more than once. 
These multi-epoch X-ray data allows us to further examine their  
long-term flux variability. We compare the difference of the fluxes with their errors combined by quadrature, i.e.:
$\left|F_{\rm obs1}-F_{\rm obs2}\right|/\sqrt{\sigma_{\rm obs1}^{1}+\sigma_{\rm obs1}^{2}}$. The largest difference found for each source are 
summarized in the column 10 in Table~4.
We consider a source to have long-term variability if the maximal difference of its flux in two observations is larger than $4\sigma$. 
Four sources, J17472\_X5, J0737\_X1, J0737\_X2 and J0737\_X5 are found to be significantly variable. For those with non-detection in 
certain epoch(s), we have placed lower bounds on their long-term variabilities instead. 

\begin{figure}
\centering
\includegraphics[width=3in,height=3in]{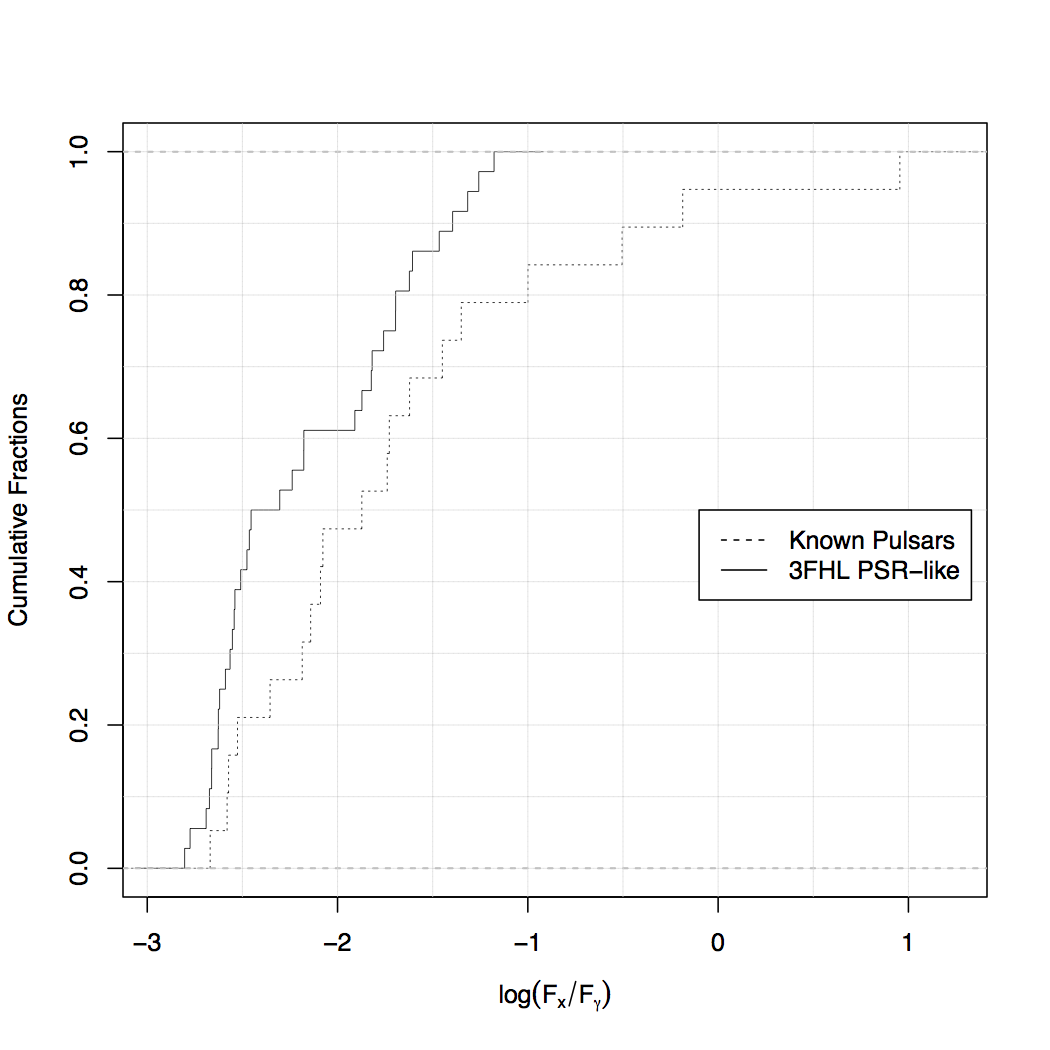}
\caption{Comparison for the distributions of $\log F_{x}/F_{\gamma}$ of known pulsars in 3FHL and the sources in Table~4. 
The fluxes in X-ray and $\gamma-$ray are evaluated in 0.3-10~keV and 10~GeV-2~TeV respectively.}
\label{cdf}
\end{figure}

Since the X-ray data provide much better constraints on the positions of the potential counterparts, we are able to 
search for the possible optical/infra-red counterparts of these X-ray sources. 
We searched the following optical and (near-)infrared source catalogs for the counterpart to the X-ray sources (search radius = 1$\arcsec$): Pan-STARRS DR2 (PS1; \citealt{2016arXiv161205560C}), GLIMPSE \citep{2009yCat.2293....0S}, the Spitzer point-source catalog of seven nearby galaxies \citep{2015ApJS..219...42K}, VISTA Variables in the Via Lactea (VVV; \citealt{2012A&A...537A.107S,2017yCat.2348....0M}), WISE all-sky catalog \citep{2012yCat.2311....0C}, and Gaia DR2 \citep{2018A&A...616A...1G}. Most of the 3FHL sources are close to the Galactic plane with heavy extinction, and several (near-)infrared catalogs were thus used. For every optical counterpart, we made an extinction corrected spectral energy distribution (SED) and fit it with the blackbody model. Dereddening was done using the extinction curve of \cite{1999PASP..111...63F} with an extinction value ($A_v$) inferred from the hydrogen column density in the X-ray analysis (i.e., $N_\mathrm{H}/Av=2.21\times10^{21}$; \citealt{2009MNRAS.400.2050G}).

Table \ref{tab:sed} shows the SED fitting results. Except for those have \textit{Gaia} distance measurements \citep{2018AJ....156...58B}, a distance of 1~kpc is assumed for the calculations of the blackbody radii (the radius is proportional to the distance). As mentioned, most of the sources are highly absorbed and therefore their SEDs are largely affected by the dereddening. Given that the extinctions adopted are full Galactic values, the SEDs of some nearby sources (e.g., J16263\_X2) could be over-corrected and appear to be much bluer than they should be.
Under-correction is also possible if a source is with high intrinsic absorption.

\begin{table*}
\begin{center}
\begin{tabular}{lccccccc}
\hline
Name & Temperature & Radius & Distance$^{\rm a}$ & Reference Magnitude$^{\rm b}$ & Extinction & offset \\
 & (K) & ($R_{\odot}$) & (kpc) & & ($A_{v}$) & (arcsec) \\
\hline
J17486\_X1 & $11800\pm1700$ & $1.3\pm0.2$ & (1) & 16.9 (0.62$\rm{\mu m}$) & 5.9 & 0.8 \\
J18007\_X2 & $1400\pm100$ & $10\pm2$ & (1) & 17.6 (1.0$\rm{\mu m}$) & 5.3 & 0.5 \\
J18007\_X3 & $3000\pm300$ & $3.1\pm0.4$ & (1) & 13.0 (2.2$\rm{\mu m}$) & 5.3 & 0.9 \\
J18007\_X5 & $21100\pm7600$ & $0.20\pm0.05$ & 0.9 & 20.1 (0.49$\rm{\mu m}$) & 5.3 & 0.4 \\
J18007\_X6 & $1300\pm200$ & $5.6\pm1.6$ & (1) & 17.1 (1.6$\rm{\mu m}$) & 5.3 & 0.6 \\
J18007\_X8 & $4700\pm200$ & $2.4\pm0.2$ & 2.5 & 19.8 (0.62$\rm{\mu m}$) & 5.3 & 0.3 \\
J14051\_X1 & $2031\pm155$ & $6.6\pm0.7$ & (1) & 14.4 (1.7$\rm{\mu m}$) & 8.5 & 0.5 \\
J14051\_X1$^{\rm c}$ & 40000 (fixed) & $12.9\pm0.8$ & 7.7 & 14.4 (1.7$\rm{\mu m}$) & 31.6 & 0.5 \\
J16263\_X1$^{\rm d}$ & - & 1.1 & (1) & 19.1 (1.6$\rm{\mu m}$) & 8.9 & 0.6 \\
J16263\_X2$^{\rm d}$ & - & - & 3.6 & 19.0 (0.51$\rm{\mu m}$) & 8.9 & 0.4 \\
J17472\_X8 & $1900\pm200$ & $11\pm2$ & (1) & 13.3 (1.7$\rm{\mu m}$) & 5.7 & 0.4 \\
J07375\_X1 & $800\pm300$ & $4.0\pm2.6$ & (1) & 15.8 (3.6$\rm{\mu m}$) & 0.20 & 0.4 \\
J07375\_X2 & $900\pm100$ & $1.3\pm0.2$ & (1) & 17.4 (3.6$\rm{\mu m}$) & 0.20 & 0.1 \\
J07375\_X3 & $5800\pm200$ & $0.18\pm0.01$ & 1.6 & 18.9 (0.49$\rm{\mu m}$) & 0.18 & 0.6 \\
J07375\_X4 & $1000\pm300$ & $0.60\pm0.33$ & (1) & 18.7 (3.6$\rm{\mu m}$) & 0.19 & 0.7 \\
J07375\_X6 & $4000\pm600$ & $0.15\pm0.04$ & (1) & 20.9 (0.62$\rm{\mu m}$) & 0.23 & 0.4 \\
J07375\_X8 & $2300\pm100$ & $0.78\pm0.05$ & (1) & 21.4 (0.62$\rm{\mu m}$) & 0.19 & 0.3 \\
\hline
\end{tabular}
\end{center}
(a) Unless a $Gaia$ distance is found, $d=1$~kpc is assumed.\\
(b) Observed magnitudes of the shortest wavelength (in the brackets) that can be found in the aforementioned catalogs. \\
(c) Distance and the extinction are adopted from Corbet et al. (2019).\\
(d) Only two data points in the SED and therefore no uncertainty can be obtained.
\caption{Results of blackbody fits to the optical/IR SED of the possible counterparts associated with 
the X-ray sources found in the error ellipses of PSR-like 3FHL sources.}
\label{tab:sed}
\end{table*}

\begin{table*}
\begin{center}
\caption{X-ray spectral properties of X-ray sources with more than fifty net counts collected from the archival data. The results from both power-law fits and 
blackbody fits are summarized. The quoted uncertainties are $1\sigma$ for one parameter of interest.}
\footnotesize
\begin{tabular}{l | c c c c c c c c c}
\hline
\hline
& \multicolumn{4}{|c|}{Power-law fit} & & \multicolumn{4}{|c|}{Blackbody fit} \\
Name & $n_H$ & $\Gamma_{x}$ & $\chi^{2}$/d.o.f & $F^{\rm unabs}_{\rm 0.3-10keV}$ & & $n_H$ & $kT$ & $\chi^{2}$/d.o.f & $F^{\rm unabs}_{\rm 0.3-10keV}$ \\
  & ($10^{22}$cm) & & & (erg cm$^{-2}$ s$^{-1}$) & & ($10^{22}$cm) & (keV) & & (erg cm$^{-2}$ s$^{-1}$)\\
\hline
J17486\_X1 & 1.3$^{+0.7}_{-0.5}$ & 3.1$^{+0.9}_{-0.7}$ & 20.42/32 & 2.1$\pm0.2\times10^{-13}$ & & 0.23$^{+0.35}_{-0.22}$ & 0.61$^{+0.12}_{-0.11}$ & 19.43/32 & 2.7$\pm0.3\times10^{-14}$ \\
& & & & & & \\
J18394\_X1 & $\leq$2.6 & 0.02$^{+0.77}_{-0.54}$ & 5.53/10 & 6.4$\pm1.0\times10^{-14}$ & & $\leq$1.8 & 2.6$^{+2.9}_{-0.9}$ & 5.80/10 & 5.3$\pm0.8\times10^{-14}$ \\
& & & & & & \\
J18233\_X1 & 1.9$^{+0.8}_{-0.6}$ & 1.1$\pm0.3$ & 38.55/40 & $2.8^{+0.9}_{-0.5}\times10^{-13}$ & & $0.44^{+0.37}_{-0.27}$ & $1.8\pm0.2$ & 38.33/40 & $2.0\pm0.2\times10^{-13}$ \\
& & & & & & \\
J18570\_X1 & 3.0$^{+1.8}_{-1.3}$ & 1.3$^{+0.7}_{-0.6}$ & 15.80/23 & 5.8$\pm0.7\times10^{-14}$ & & 1.3$^{+1.1}_{-0.7}$ & 1.6$^{+0.6}_{-0.4}$ & 17.91/23 & 3.6$\pm0.4\times10^{-14}$ \\
& & & & & & \\
J14051\_X1 & 15.0$^{+8.0}_{-5.0}$ & 2.7$^{+1.4}_{-1.1}$ & 13.97/21 & 2.4$^{+53.2}_{-1.9}\times10^{-12}$ & & 9.2$^{+5.1}_{-3.4}$ & 1.3$^{+0.4}_{-0.3}$ & 13.71/21 & 3.0$^{+1.4}_{-0.7}\times10^{-13}$ \\
& & & & & & \\
J16263\_X1 & 0.65$^{+1.03}_{-0.65}$ & 2.0$^{+0.7}_{-0.6}$ & 16.31/23 & 2.7$^{+3.8}_{-0.9}\times10^{-13}$ & & $\le$0.34 & 0.90$\pm0.10$ & 15.78/23 & 1.2$\pm0.2\times10^{-13}$ \\
& & & & & & \\
J16263\_X3 & 0.84$^{+0.24}_{-0.20}$ & 4.0$^{+0.7}_{-0.6}$ & 10.51/23 & 1.9$\pm0.2\times10^{-12}$ & & 0.16$^{+0.19}_{-0.15}$ & 0.40$^{+0.07}_{-0.06}$ & 13.16/23 & 1.2$\pm0.1\times10^{-13}$ \\
\hline
\end{tabular}
\end{center}
\end{table*}

\subsection{Detailed Analysis of Individual PSR-like Candidates}
The details of the X-ray observations and data analyses of these ten PSR candidates are given in the followings:
\subsubsection{3FHL J1748.6-2816}
Both {\it Chandra} (Obs ID: 2269) and {\it XMM-Newton} (Obs ID: 0694641401) have observed the field of 3FHL J1748.6-2816 on 2001 July 16
and 2012 September 30 for the effective exposures of 18~ks and 32~ks respectively. In both observations, only one X-ray source is detected 
within the $\gamma-$ray positional error ellipse which is denoted as J17486\_X1 (see Figure~4). Searching in SIMBAD, we found the nature of this 
source remains to be unidentified. For estimating its absorption-corrected 
X-ray flux as given in Table~4, we adopted the count rate from the {\it Chandra} observation and assumed 
a column absorption of $n_{H}=1.3\times10^{22}$~cm$^{-2}$ at the same level as the Galactic HI 
column density in the corresponding direction \citep{kalberla2005}.

Besides J17486\_X1, there are other sources are detected serendipitously in the whole field-of-view (FoV) 
covered by the cameras in both observations. 
We have considered the possibility that one or more sources lie within the error ellipse by chance. 
We counted the number of X-ray sources detected in the whole FoV and computed the source density. Based on this, we estimated
the number of chance coincidences $\lambda$ expected within the $\gamma$-ray error ellipse. Assuming a Poisson distribution,
the probability of finding one or more chance coincidences of X-ray sources is given by:

\begin{equation}
P\left(n\geq 1\right)=\sum_{n=1}^{\infty}\frac{\lambda^{n}e^{-\lambda}}{n!}=1-e^{-\lambda}
\end{equation}

For 3FHL~J1748.6-2816, we found that $P\left(n\geq 1\right)\sim40\%$ and $\sim34\%$ in {\it Chandra} and {\it XMM-Newton} observations 
respectively. J17846\_X1 does not show any X-ray flux variability neither in individual observations nor between two observations at different epoch.
Optical/IR counterpart of J17846\_X1 has been identified. A blackbody fit to its extinction-corrected SED yields a temperature 
of $T_{\rm bb}\sim1.2\times10^{4}$~K and an emitting region with a radius of $R_{\rm bb}\sim1.3d_{\rm kpc}$~$R_{\odot}$ (cf. Table~5), where
$d_{\rm kpc}$ is the distance at unit of 1~kpc.

In this work, a detailed X-ray spectral fitting will be carried out for those sources with more than 50 net counts detected.
The results are summarized in Table~6. 
Since the net counts of J17486\_X1 collected from both observations is $\sim140$~cts, we have extracted its spectrum and fitted with 
both absorbed power-law model and absorbed blackbody model. Both models result in a similar goodness-of-fit. 
The best-fit power-law yields a column absorption of 
$n_{H}=1.3^{+0.7}_{-0.5}\times10^{22}$~cm$^{-2}$, a photon index of $\Gamma_{x}=3.1^{+0.9}_{-0.7}$ and an absorption-corrected of 
$F_{x}\sim2.1\times10^{-13}$~erg~cm$^{-2}$~s$^{-1}$ in 0.3-10~keV. The best-fitted $\Gamma\sim3$ appears to be quite steep which indicate the 
X-ray emission is rather soft. Considering a purely thermal emission scenario, the best-fit blackbody yields a temperature of
$kT=0.6\pm0.1$~keV with an absorption-corrected of $F_{x}\sim2.7\times10^{-14}$~erg~cm$^{-2}$~s$^{-1}$ in 0.3-10~keV. The normalization of the blackbody
implies an X-ray emission region with a radius of $\sim13.4d_{\rm kpc}$~m. 

\subsubsection{3FHL J1839.4-0553}
The $\gamma-$ray error ellipse of 3FHL J1839.4-0553 has been covered by two {\it Chandra} observations with ACIS-I on 2008 March 9 (Obs. ID: 7493)
and 2007 November 5 (Obs. ID. 7630) for an effective exposure of 20~ks and 28~ks respectively. 
Within the 95\% $\gamma-$ray error ellipse, there are two X-ray sources 
J18394\_X1 and J18394\_X2 (see Figure~4). 
J18394\_X1 has been detected by both observations. Therefore, we are able to estimate its long-term variability which is only at 
$1.4\sigma$ level. On the other hand, J18394\_X2 is out of the FoV in one {\it Chandra} observation (Obs. ID. 7630). 
The nature of both X-ray sources is not known. 
Their absorption-corrected X-ray fluxes as given in Table~4 is estimated by assuming a column 
absorption of $n_{H}=1.8\times10^{22}$~cm$^{-2}$, which is consistent with the total Galactic HI absorption at that direction, 
with the count rates observed by Obs. ID: 7493. Taking all the X-ray sources detected in the entire FoV in the observation, $P\left(n\geq 1\right)$ 
is found to be $\sim65\%$. Searching for their 
counterparts in other wavelengths with the archival data does not yield any positive result. 

J18394\_X1 has $\sim58$ net counts collected from both observations and therefore we have further examined its X-ray spectrum (see Table~6). 
Its X-ray spectrum appears to be rather flat in the energy range of 0.5-8~keV. Both power-law and blackbody fits suggest the column absorption can be lower 
than that inferred from the Galactic HI absorption. The best-fit power-law yields $n_{H}<2.6\times10^{22}$~cm$^{-2}$, $\Gamma_{x}=0.02^{+0.77}_{-0.54}$ and 
$F_{x}\sim6.4\times10^{-14}$~erg~cm$^{-2}$~s$^{-1}$ in 0.3-10~keV. On the other hand, the best-fit blackbody yields 
$n_{H}<1.8\times10^{22}$~cm$^{-2}$, $kT=2.6^{+2.9}_{-0.9}$~keV and $F_{x}\sim5.3\times10^{-14}$~erg~cm$^{-2}$~s$^{-1}$ in 0.3-10~keV.

\begin{figure*}
\centering
\includegraphics[width=1.0\textwidth]{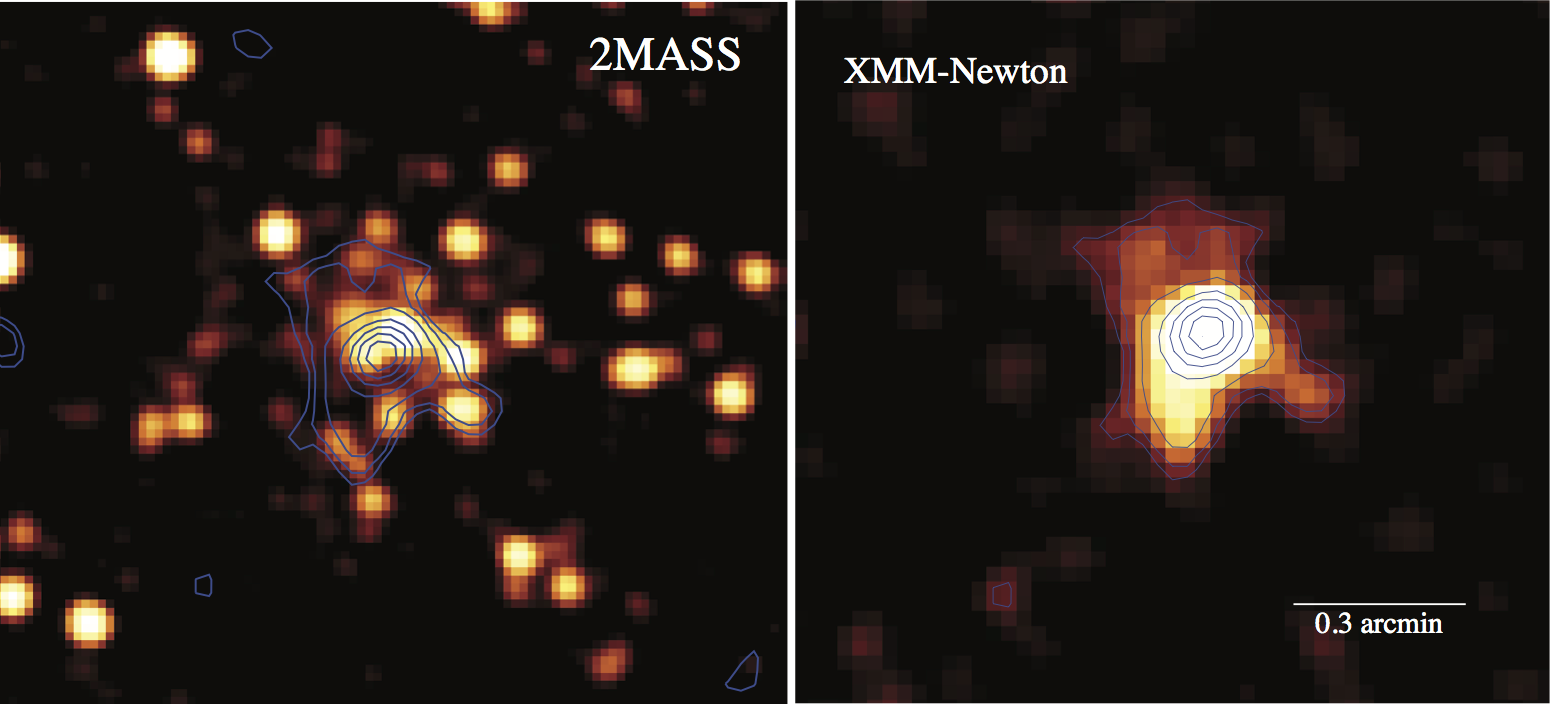}
\caption{{\it Left panel:} $K_{s}$ band image of a field centered at
Mercer 5 obtained by 2MASS. A central concentration of stars can be noted.
{\it Right panel:} X-ray image of J18233\_X1 in 0.3-10~keV with data from MOS1 and MOS2
onboard {\it XMM-Newton} combined. This is the only X-ray source lies within the $\gamma-$ray 
error ellipse of 3FHL J1823.3-1339. An apparently extended X-ray feature
is discovered at the location of Mercer~5. We overlay the X-ray contours on
the infrared image for comparing the morphology at different wavelengths. Top is north and left is east.}
\label{mercer5_x_ir}
\end{figure*}

\subsubsection{3FHL J1823.3-1339}
3FHL J1823.3-1339 has been observed by {\it XMM-Newton} on 23 March 2002 (Obs. ID: 0040140201) for an effective exposure of 
$\sim11$~ks. Only one single X-ray source, J18233\_X1, is detected within its $\gamma-$ray error ellipse. Its X-ray image as 
observed by MOS1/2 camera onboard {\it XMM-Newton} is displayed in Figure~4. 
Using the serendipitous X-ray sources detected in the FoV of MOS camera, the probability of finding an X-ray source within the
error ellipse of 3FHL J1823.3-1339 is found to be $P\left(n\geq 1\right)\sim12\%$, which is the lowest among
all the selected 3FHL PSR-like candidates in this work.
We noticed that the feature is apparently extended with an angular size of $\sim30$~arcsec and has the peak emission located at 
RA~(J2000)=18$^{\rm h}$23$^{\rm m}$19$^{\rm s}$ Dec~(J2000)=$-13^{\circ}40^{'}02^{''}$. 
This extended X-ray feature is identified for the first time. A close-up view of J18233\_X1 is shown in the right panel of Figure~6.

Searching for the nature of this extended source in SIMBAD, we found that it is possibly associated with a poorly-studied globular cluster 
Mercer 5, which is discovered in the GLIMPSE Survey \citep{mercer2005}. It is highly obscured in optical regime as it resides in a region of high
visual extinction, $A_{V}\sim8.5-12.5$~mag \citep{longmore2011}, which suggests an X-ray absorption at the level of 
$\left(1.2-2.8\right)\times10^{22}$~cm$^{-2}$ \citep{2009MNRAS.400.2050G}.
In left panel of Figure~6, we compare the X-ray morphology of J18233\_X1 with the $K_{s}$ band 2MASS image of Mercer 5. 
by overlaying the X-ray contours on the infrared image. The distribution of the stars in Mercer~5 is comparable
with the morphology of J18233\_X1. The peak of the X-ray emission coincides with the region with highest stellar density.

The net counts of J18233\_X1 collected from all EPIC cameras on {\it XMM-Newton} (MOS1/2 + PN) is 322~cts. 
This enables us to carry out a detailed analysis. 
In examining its X-ray spectrum, we found that it can be well-described by an absorbed power-law model with a goodness-of-fit of 
$\chi^{2}=38.55$ for 40 d.o.f.. The observed X-ray spectra of J18233\_X1 and the best-fitted power-law model are displayed in Figure~7. 
The X-ray emission of J18233\_X1 is quite hard. 
The best-fit yields
a column absorption of $n_{H}=1.9^{+0.8}_{-0.6}\times10^{22}$~cm$^{-2}$,  
a photon index of  $\Gamma_{x}=1.1\pm0.3$ and 
an absorption-corrected
flux in 0.3-10~keV of $F_{x}\sim3\times10^{-13}$~erg~cm$^{-2}$~s$^{-1}$. The X-ray column absorption inferred from the spectral fit is consistent 
with that deduced from the $n_{H}-A_{v}$ correlation. This suggests that J18233\_X1 and Mercer~5 are very likely to be located at the same distance from us.  

We have also attempted to search for X-ray periodicity from J18233\_X1. However, we do not find any significant periodic signal from the existing data. 

\begin{figure}
\begin{center}
\includegraphics[angle=-90,width=8.5cm]{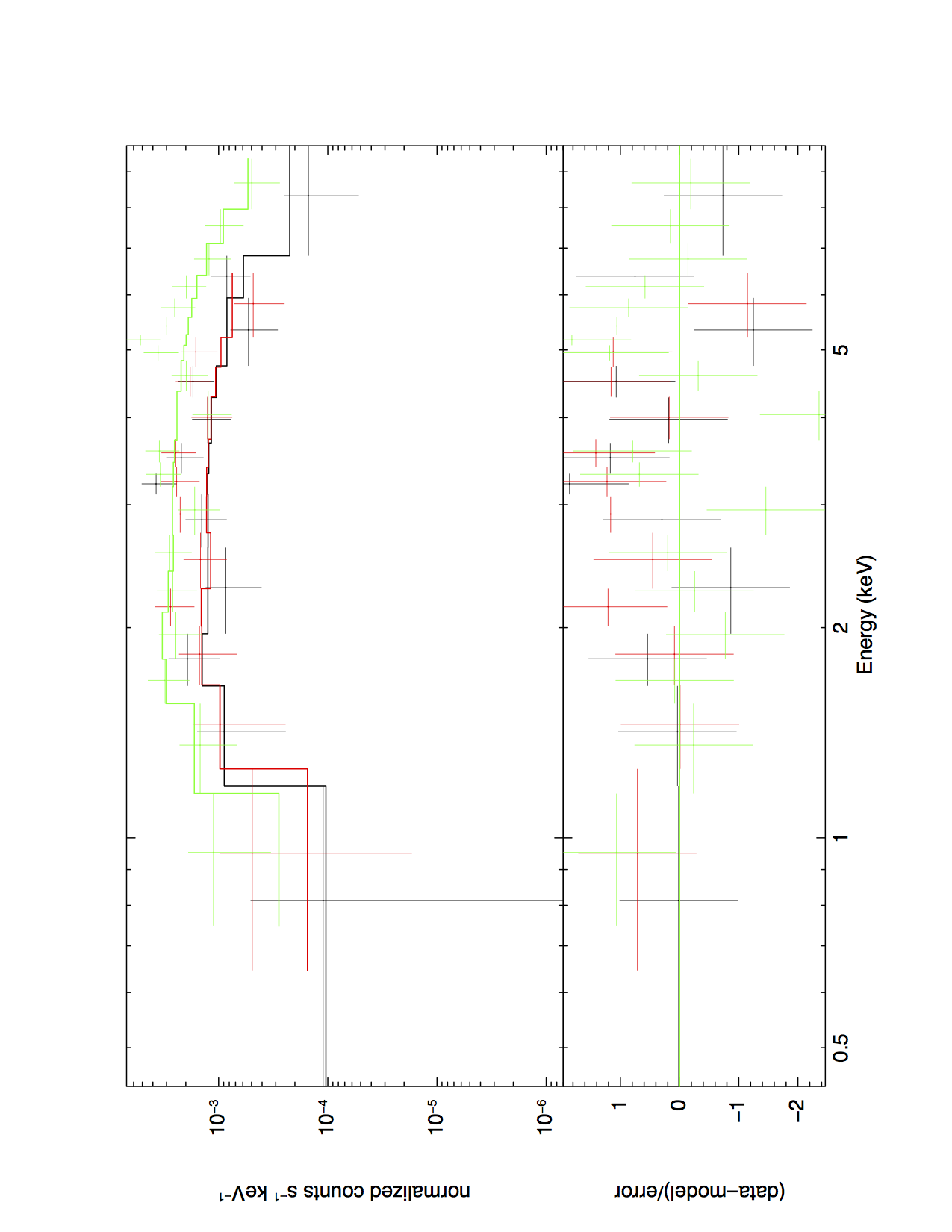}
\end{center}
\caption{The X-ray spectra of J18233\_X1 which is positionally coincident with 
the globular cluster Mercer~5 as observed by \emph{XMM-Newton} MOS1/2 + PN cameras and
simultaneously fitted to an absorbed power-law(upper panel)
and contribution to the fitting residuals (lower panel).}
\end{figure}

We further investigated if 3FHL~J1823.3-1339 can also be the
$\gamma-$ray counterpart of Mercer~5. 3FHL~J1823.3-1339 is also
identified in the 3FGL catalog with designation 3FGL~1823.2-1339 \citep{acero2015}.
In 0.1-100 GeV, its energy flux is
$f_{\gamma}=(9.3\pm0.8)\times10^{-11}$~erg~cm$^{-2}$~s$^{-1}$. At a distance
of $d\sim5.5$~kpc as estimated by {\it Gaia} (DR2) \citep{baumgardt2019},
this corresponds to a $\gamma-$ray luminosity of $L_{\gamma}\sim3.4\times10^{35}$~erg/s.
On the other hand, the metallicity of Mercer~5 is estimated to be [Fe/H]$\sim$-0.86
\citep{penaloza2015}.
Using its metalicity, we can estimate the expected $L_{\gamma}$ from a globular cluster 
by using the empirical relation: 
$\log L_{\gamma}=(0.6\pm0.2)$[Fe/H]+(35.6$\pm$0.2) \citep{hui2011}.
This implies that the $\gamma-$ray luminosity of Mercer~5 is expected at the order of $\sim10^{35}$~erg/s.
This is consistent with the observed luminosity within the tolerance of the uncertainties 
of fitted parameters and $L_{\gamma}$. This suggests the $\gamma-$rays from 
3FHL~J1823.3-1339/3FGL~1823.2-1339 are likely from Mercer~5. 

\subsubsection{3FHL J1748.1-2903}
3FHL J1748.1-2903 has been observed by two {\it Chandra} observations with ACIS-I CCD array  
on 2006 October 31 (Obs. ID. 7158) and 2017 July 13 (Obs. ID. 19448) with effective exposures of $\sim14$~ks and $\sim45$~ks respectively. 
Using these data, two sources namely J17481\_X1 and J17481\_X2 are detected within the $\gamma-$ray error ellipse. A merged image is shown in 
Figure~4. However, J17481\_X1 can only be detected in the 2017 observation and J17481\_X2 can only be detected in the 2006 observation. 
Based on the limiting flux in the corresponding epoch of non-detection, we placed the limits on the long-term variability for J17481\_X1 and J17481\_X2
as $>2\sigma$ and $>3\sigma$ respectively. Both of them are potentially variable X-ray sources. 
The absorption-corrected X-ray fluxes of J17481\_X1 J17481\_X2 tabulated in Table~4 is estimated with their count rates in the corresponding observation 
and a total Galactic HI column density of $n_{H}=1.1\times10^{22}$~cm$^{-2}$. No optical/IR counterpart were found for these two sources. 
The net counts for both sources are $<50$~cts and therefore no further analysis will be proceeded. 
 
\subsubsection{3FHL J1857.0+0059}
3FHL J1857.0+0059 has been observed by {\it XMM-Newton} (Obs. ID. 0784040201) on 2016 October 13 for an effective exposure of $\sim37$~ks. 
Within its $\gamma-$ray ellipse, only one X-ray source J18570\_X1 is detected in this data (cf. Figure 4). The nature of J18570\_X1 remains unidentified in SIMBAD. 
The $P\left(n\geq 1\right)$ inferred from this observation is $\sim40\%$. The absorption-corrected X-ray flux of J1857\_X1 as given in Table~4 
is estimated by assuming a total Galactic HI column density of $n_{H}=1.1\times10^{22}$~cm$^{-2}$. Searching for its multiwavelength counterpart does not yield any result.

On the other hand, we noted that a pulsar PSR J1857+0057 is lying within the 3FHL error circle. The angular separation between PSR J1857+0057 and 
J18570\_X1 is $\sim5.5$~arcmin. Therefore, there is no association between these two objects. PSR J1857+0057 has a spin-down power of 
$\dot{E}=4.7\times10^{31}$~erg/s \citep{manchester2005}. At a distance of $d\sim2.5$~kpc as inferred by the dispersion 
measure of this pulsar, 3FHL J1857.0+0059/3FGL~J1857.2+0059 has a luminosity of $L_{\gamma}\sim3.6\times10^{34}$~erg/s at 
energies $>100$~MeV which is three orders 
of magnitude larger than $\dot{E}$. Therefore, we concluded that PSR J1857+0057 cannot be associated with this $\gamma-$ray source. 

There are $\sim85$ net counts collected from J18570\_X1 in this {\it XMM-Newton} observation. In examining its X-ray spectrum, we found 
that a best-fit power-law yields $n_{H}=3.0^{+1.8}_{-1.3}\times10^{22}$~cm$^{-2}$, $\Gamma_{x}=1.3^{+0.7}_{-0.6}$ and 
$F_{x}\sim5.8\times10^{-14}$~erg~cm$^{-2}$~s$^{-1}$ in 0.3-10~keV. And a best-fit blackbody yields
$n_{H}=1.3^{+1.1}_{-0.7}\times10^{22}$~cm$^{-2}$, $kT=1.6^{+0.6}_{-0.4}$~keV and $F_{x}\sim3.6\times10^{-14}$~erg~cm$^{-2}$~s$^{-1}$ in 0.3-10~keV.

\begin{figure}
\includegraphics[width=8.5cm]{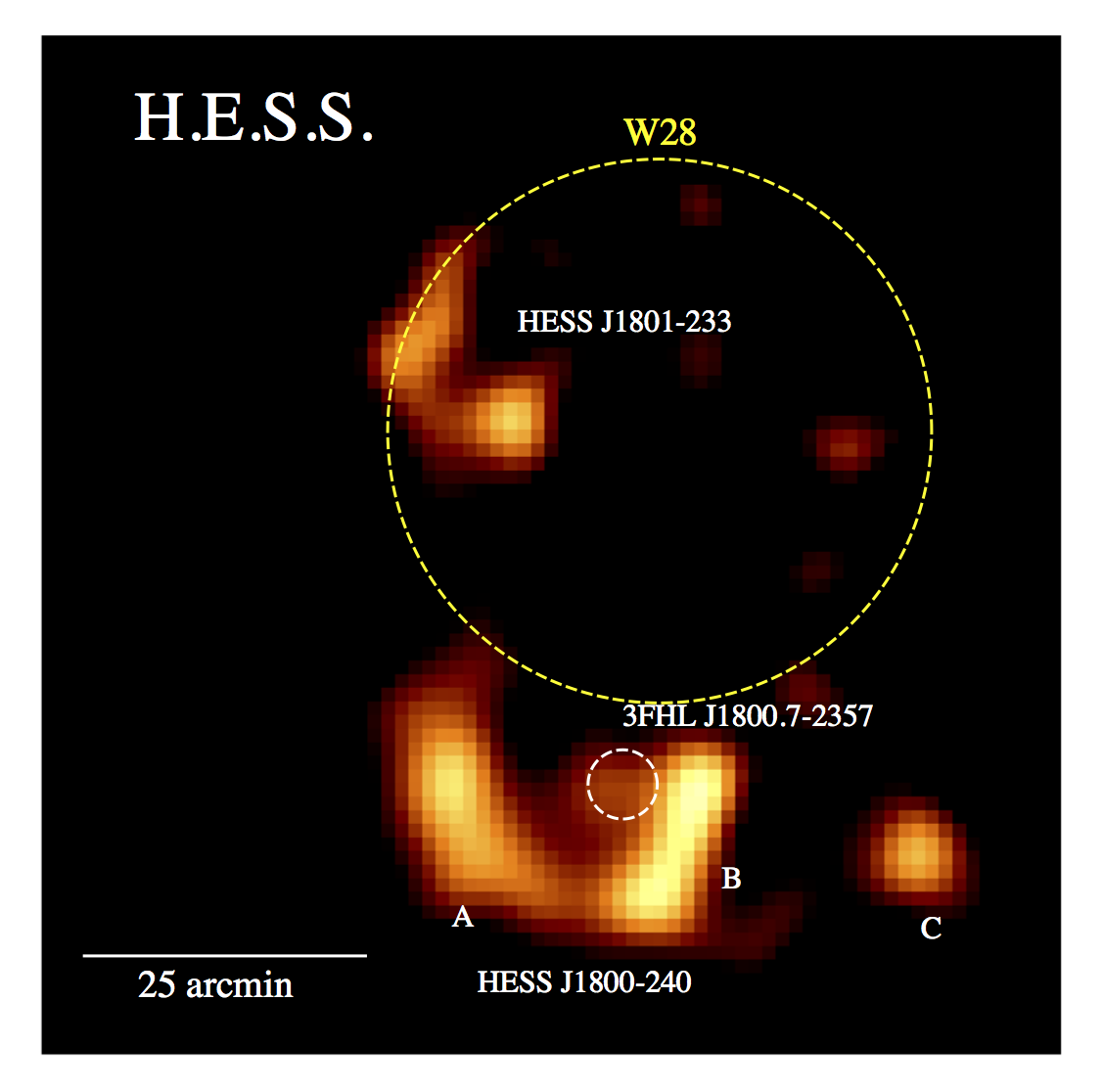} 
\caption{A $\gamma-$ray excess map at energies $>0.1$~TeV of the W28 field as obtained by H.E.S.S. \citep{aharonian2008}. 
The VHE sources HESS~J1801-233 and the complex 
of HESS~J1800-240 (regions A, B \& C) can be clearly seen. The location and the angular size of the supernova remnant W28 is illustrated by 
the dotted yellow circle. The dotted white ellipse is the 95\% confidence $\gamma-$ray positional uncertainty of 3FHL J1800.7-2357.}
\end{figure}

\subsubsection{3FHL J1800.7-2357}
3FHL J1800.7-2357 has been observed by {\it Chandra} (Obs. ID. 10997) on 2010 July 30 for an effective exposure of $\sim80$~ks. 
3FHL J1800.7-2357/3FGL J1800.8-2402 resides in a region where the supernova remnant (SNR) W28 interacts with a number of surrounding 
molecular cloud (MC) \citep{aharonian2008}. A complex of TeV emission is found in this field (see Figure~8). 3FHL J1800.7-2357 apparently 
coincides with a protrusion at the eastern edge of HESS J1800-240B. This leads us to speculate whether this feature is indeed a distinct source 
or a part of the $\gamma-$rays from the SNR-MC interactions.
For investigating this issue, VHE observation facility with improved spatial resolution is required (e.g. CTA). 

\cite{2018ApJ...860...69C} have reported an updated analysis of this field with 9 years {\it Fermi} LAT data. In a 
10-200 GeV sky map, they have found features which spatially match HESSJ1800-240A, HESSJ1800-240B and HESSJ1800-240C. 
HESSJ1800-240B is the brightest among them \cite[Figure~1 in][]{2018ApJ...860...69C}. The GeV spectra of both 
HESSJ1800-240B and HESSJ1800-240C show flux discontinuities which suggests there can be several emission components contribute 
to the $\gamma-$rays detected at their locations. While the emission below $\sim1$~GeV can come from the a nearby source with 
unknown origin, \cite{2018ApJ...860...69C} argue that the $\gamma-$rays with energies $\gtrsim1$~GeV 
from all three spatial components of HESSJ1800-240 
have a hadronic origin which is dominated by the interactions with the local sea of Galactic cosmic rays.

On the other hand, HESSJ1800-240B is potentially associated with a massive star formation region G5.89-0.39 \citep{hampton2016}. This suggests finding 
young neutron stars or pulsars in this region is not unreasonable. However, the high density of X-ray sources in this region 
makes the probability of having more than one chance coincidence within the error ellipse of 3FHL J1800.7-2357, $P\left(n\geq 1\right)$, almost close to 100\%. 
Eight X-ray sources are detected within the $\gamma-$ray error ellipse (Figure 4). Total Galactic HI column density of $n_{H}=1.2\times10^{22}$~cm$^{-2}$ 
and the count rates of these sources obtained in this observation are adopted for 
estimating their $F_{x}$ (cf. Table~4). Based on Gregory-Loredo variability algorithm, J18007\_X1 is the only X-ray source that in this investigation that shows 
possible variability in a single observation with a probability $>90\%$. Its X-ray light curve is shown in Figure~9. 
Since the net counts collected from J18007\_X1 is $<50$~cts, we do not carry out any further 
analysis of this source. 

We have also identified the optical/IR counterparts of J18007\_X2, J18007\_X3, J18007\_X5, J18007\_X6 and J18007\_X8. 
By fitting the blackbody model to their SED, temperatures in the range of $T\sim1300-21100$~K are yielded (Table~5). 
For J18007\_X5 and J18007\_X8, their counterparts can also be found in {\it Gaia} DR2. Parallax measurements suggest J18007\_X5 and J18007\_X8 
are located at the distance of 0.9~kpc and 2.5 kpc respectively. Adopting these distances, the blackbody radii of the optical/IR counterparts of 
J18007\_X5 and J18007\_X8 are found to be $0.2R_{\odot}$ and $2.4R_{\odot}$ respectively. 

\begin{figure}
\begin{center}
\includegraphics[width=9cm]{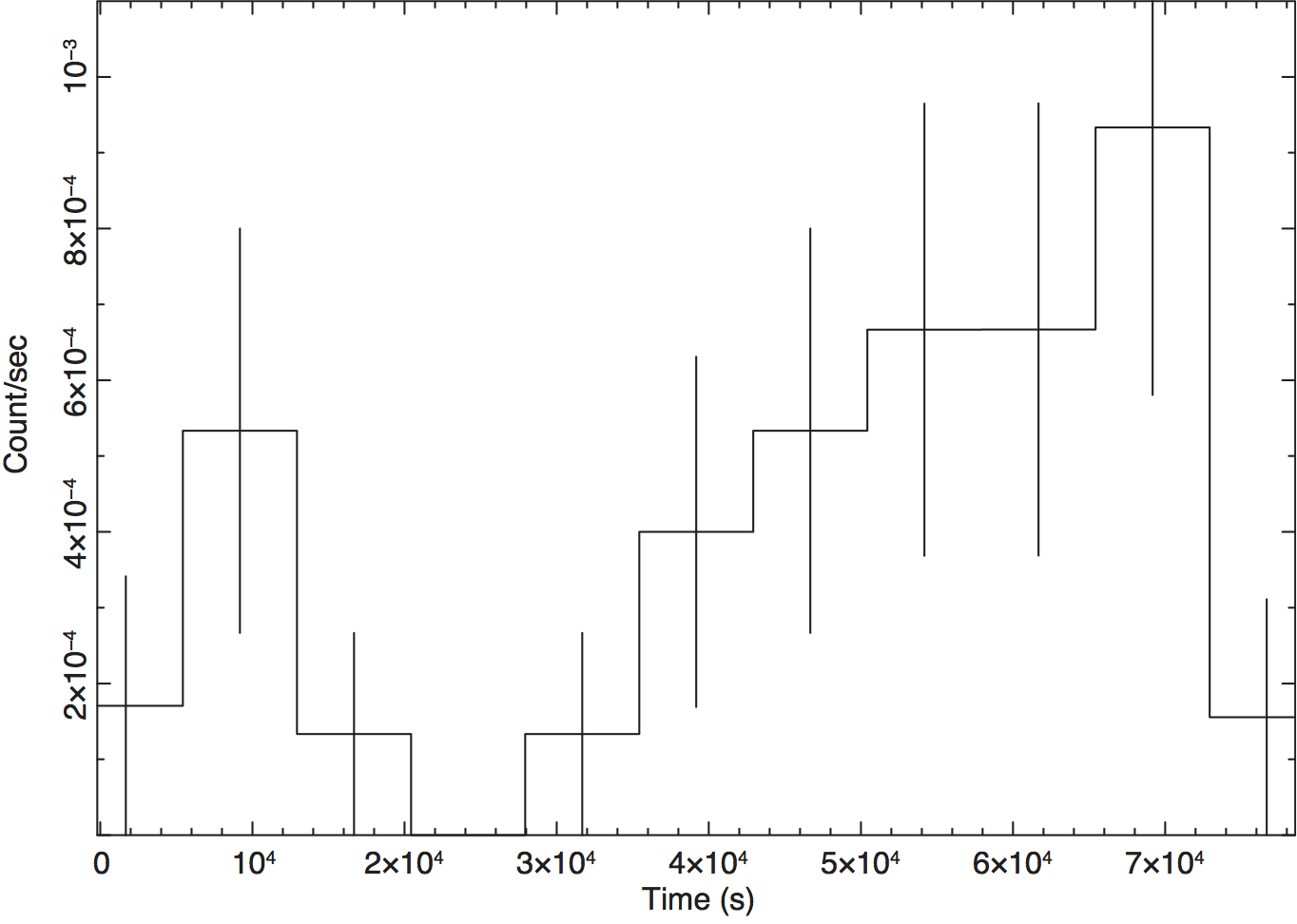}
\end{center}
\caption{The X-ray light curve of J18007\_X1 with a bin size of 7500~s as observed by {\it Chandra} in 0.3-7~keV. }
\end{figure}

\subsubsection{3FHL J1405.1-6118}
3FHL J1405.1-6118 has been observed by {\it Chandra} ACIS-S on 2013 September 19 (Obs. ID. 14888) for an effective exposure of $\sim13$~ks. 
Two X-ray sources, J14051\_X1 and J14051\_X2, are detected within the $\gamma-$ray error ellipse. 
The probability of chance coincidence is estimated to be $P\left(n\geq 1\right)\sim34\%$. Searching in SIMBAD, we found that 
both X-ray sources are unclassified. Their $F_{x}$ as given in Table~4 are estimated with their detected count rates and 
a total Galactic HI column density of 
$n_{H}=1.9\times10^{22}$~cm$^{-2}$. 
J14051\_X1 is among the brightest X-ray sources discovered in this work, which is detected at a S/N ratio of $\sim24\sigma$. 
$\sim69$ net counts from this source have been collected by ACIS-S in this observation and this allows us to perform a detailed analysis. 

\begin{figure}
\begin{center}
\includegraphics[angle=-90,width=9cm]{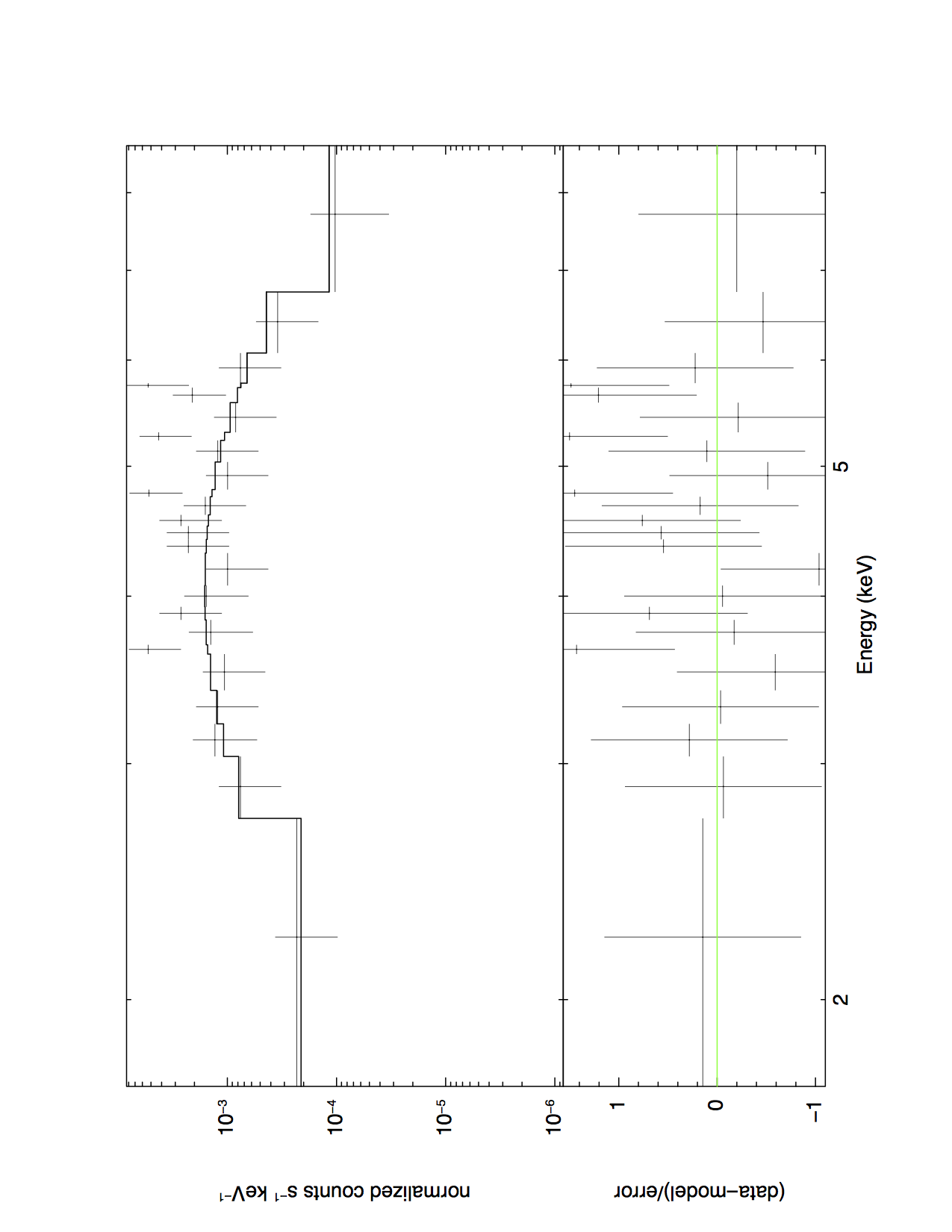}
\end{center}
\caption{The X-ray spectrum of J14051\_X1 
as observed by \emph{Chandra} ACIS-S with the best-fitted absorbed power-law(upper panel)
and contribution to the fitting residuals (lower panel).}
\end{figure}

\begin{figure}
\begin{center}
\includegraphics[angle=-90,width=8cm]{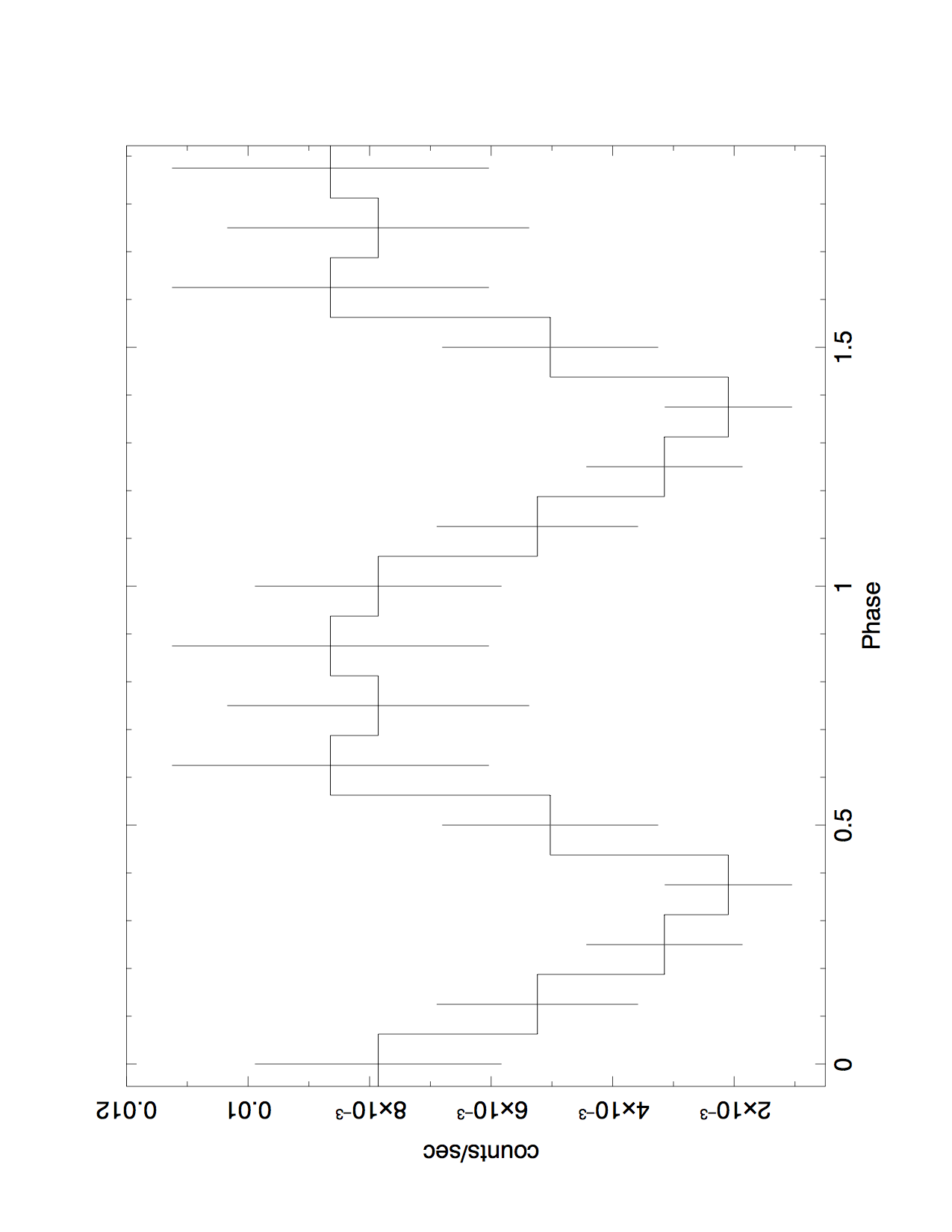}
\end{center}
\caption{X-ray light curve of J14051\_X1 in 0.5-7~keV as obtained by \emph{Chandra} ACIS-S. It is folded at the period of $P=1.4$~hrs. }
\end{figure}
We found its X-ray spectrum can be well-described by an absorbed power-law model with a photon index of $\Gamma_{x}=2.7^{+1.4}_{-1.1}$ (Figure 10). 
It yields a goodness-of-fit of $\chi^{2}=13.97$ for 21 d.o.f..
The best-fit column absorption is found to be $n_{H}=1.5^{+0.8}_{-0.5}\times10^{23}$~cm$^{-2}$ 
which is much larger than the total Galactic HI column density along the direction toward this source. Adopting this best-fit model, the absorption-corrected 
X-ray flux becomes $F_{x}=2.4\times10^{-12}$~erg~cm$^{-2}$~s$^{-1}$ in 0.3-10 keV. On the other hand, its spectrum can also be fitted by an absorbed blackbody 
model and results in a comparable goodness-of-fit ($\chi^{2}=13.71$ for 21 d.o.f.). It yields an $n_{H}=9.2^{+5.1}_{-3.4}\times10^{22}$~cm$^{-2}$ 
and a temperature of $kT=1.3^{+0.4}_{-0.3}$~keV. The best-fit normalization implies a thermal emission region 
with a radius of $\sim10.6~d_{\rm kpc}$~m where $d_{\rm kpc}$ is the distance to the source in unit of kpc. The best-fit blackbody model implies 
an absorption-corrected X-ray flux to be $F_{x}\sim3.0\times10^{-13}$~erg~cm$^{-2}$~s$^{-1}$ in 0.3-10 keV.

We have also examined its X-ray temporal properties. 
Although the variability analysis with 
the Gregory-Loredo algorithm does not indicate any significant variability, a visual inspection of its barycentric-corrected 
light curve suggest a possible underlying structure. In particular, 
there appears to have two peaks separated by $\sim5000$~s which suggests a possible periodicity. Searches for the periodic signal around this value by 
epoch-folding yield a candidate 
signal at $P\sim5090$~s with $\chi^{2}=12.2$ for 7 d.o.f.. The X-ray light curve of J14051\_X1 folded at this putative period is shown in Figure~11. 
Although the statistical significance of this folded light curve for being different from a uniform distribution is low (pre-trial $p-$value $\sim10\%$), 
its apparently sinusoidal nature makes it as a promising candidate for further investigation.

Visual examination of the X-ray image of J14051\_X1 suggests it is possibly extended. A close-up view of J14051\_X1 is shown in Figure~12. 
The source appears to be slightly elongated along the northwest-southeast orientation. Also, it apparently extends towards southwest. 
In order to further investigate its spatial nature, we compute its brightness profiles along the aforementioned orientations with sampling regions 
illustrated by the upper panels in Figure 13. 
In the lower-right panel of Figure 13, we show the brightness profile of J14051\_X1 along the northwest-southeast orientation. The source appears to have 
an extent of $\sim10$~arcsec towards northwest. As J14051\_X1 has an off-axis angle of $\sim4.1$~arcmin in this observation, the apparent elongation can 
be a result of distorted point spread function (PSF). To examine this, we have used the {\it Chandra} Ray Tracer 
({\it ChaRT}) to simulate the PSF. The adopted inputs for simulating the PSF are the energy spectrum of J14051\_X1 with the same exposure, roll, and off-axis angle
as in the ACIS-S3 observation. Then we computed the brightness profile of the simulated data with the same set of sampling regions in the upper-right
panel of Figure~13. The result is displayed as the dotted line in the lower-right panel of Figure~13, which matches the observed profile pretty well. 
Hence, we conclude that the elongation of J14051\_X1 along the northwest-southeast orientation is due to the degraded angular resolution as a result of 
large off-axis angle. 

On the other hand, the brightness profile for the southwestern extended feature is shown in the lower-left panel of Figure 13. 
The feature appears to have an extension of $\sim20$~arcsec towards southwest before it falls to the background. 
We have also compared the observed profile with the simulated PSF. In this direction, the profile of the simulated PSF falls to the background within the 
the bin corresponds to the peak in the observed profile. Therefore, this $\sim20$~arcsec extent cannot be accounted by the distorted PSF. 
The signal-to-noise ratio of this feature is $\sim4\sigma$. 
We have examined the Digitized Sky Survey optical image for the region of this feature. We 
do not find any optical counterpart to account for this putative extended X-ray feature.

\begin{figure}
\begin{center}
\includegraphics[width=3in]{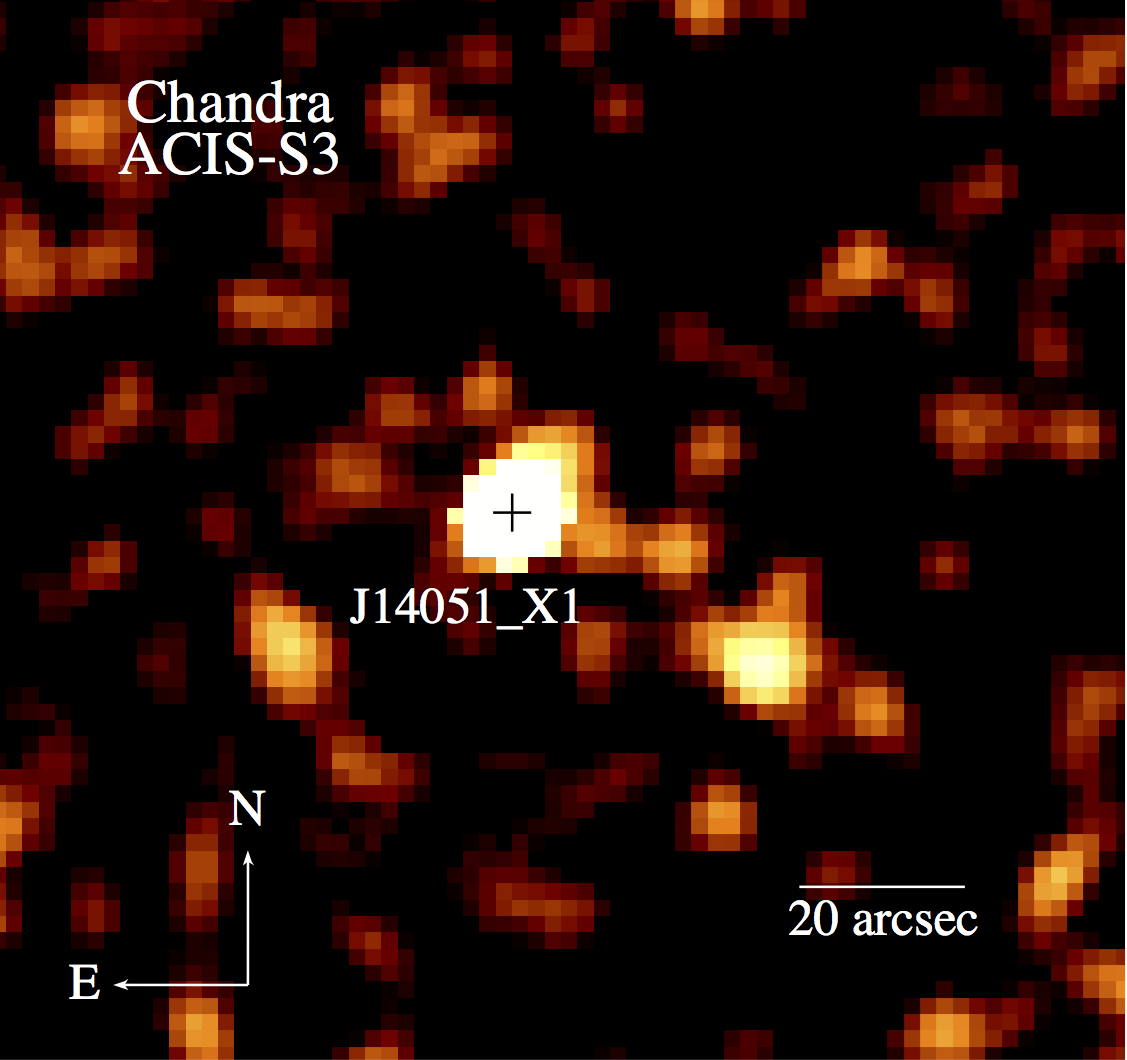}
\end{center}
\caption{A smoothed image of the field around J14051\_X1 as observed by {\it Chandra} ACIS-S3 in 0.3-8~keV. The black cross illustrates the 
X-ray position of J14051\_X1 as given in Table~4. The source is 
apparently elongated along the NW-SE orientation and it also appears to be extended towards SW.}
\end{figure}

\begin{figure*}
\begin{center}
\includegraphics[width=6in]{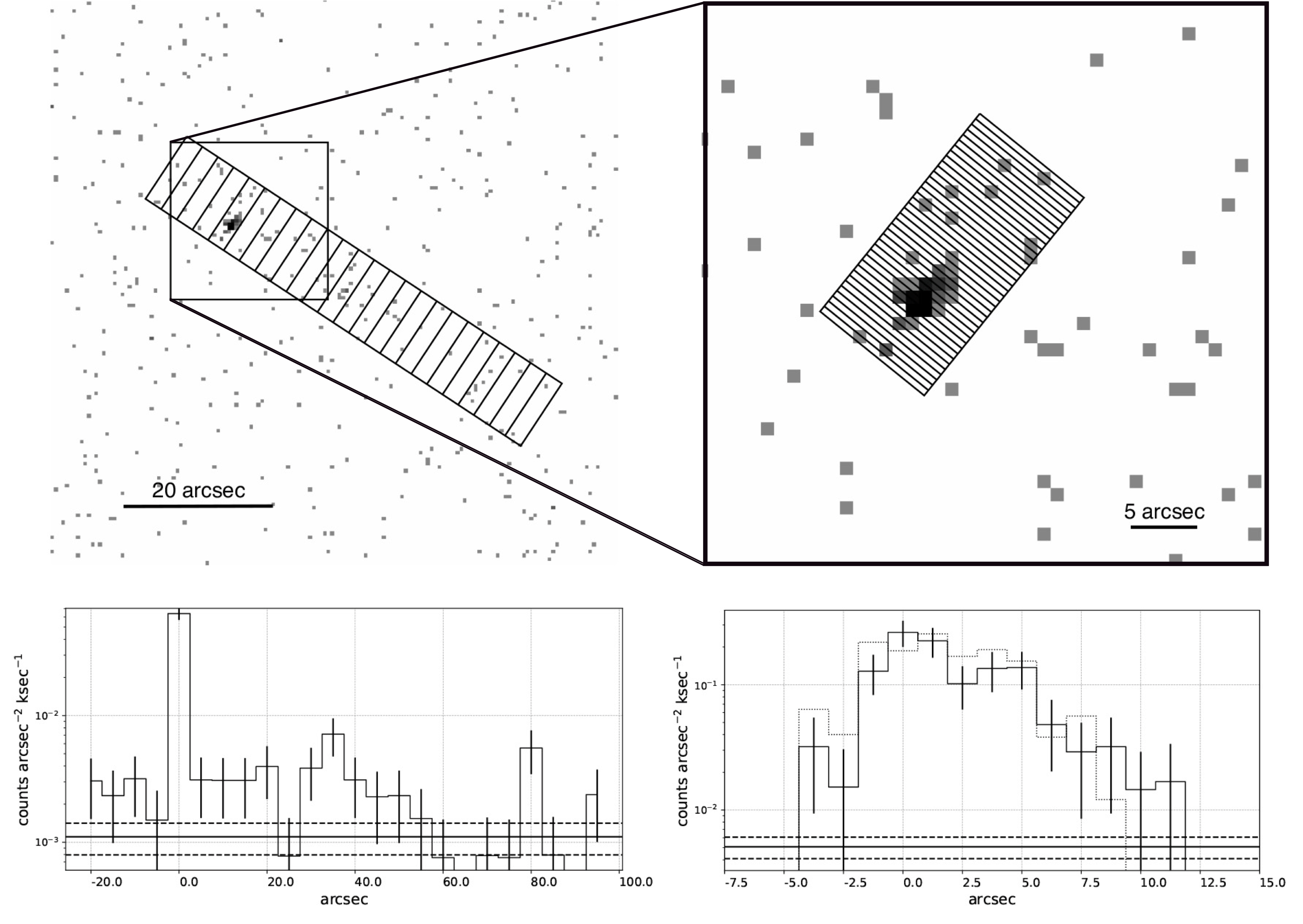}
\end{center}
\caption{The brightness profile of the putative SW extended feature associated with J14051\_X1 (lower-left panel) 
as sampled from regions in the {\it Chandra} ACIS-S3 raw image (upper-left panel). The brightness profile of J14051\_X1 along the NW-SE orientation 
as sampled from the regions illustrated in the upper-right panel is displayed in the lower-right panel. The dotted line in this plot represents the 
expected profile from a point-like source. 
The average background level and its $1\sigma$
deviation are indicated by horizontal lines that were calculated by sampling from the source-free regions.}
\label{J14051_profile}
\end{figure*}

We have also identified the IR counterpart of J14051\_X1. Using the extinction of $A_{v}=8.5$ as inferred from the Galactic HI column density, we 
constructed the extinction-corrected SED. Fitting a blackbody to this SED yields a temperature of 
$T_{\rm bb}\sim2013$~K and a emitting area with a radius of $R_{\rm bb}\sim6.6R_{\odot}d_{\rm kpc}$ with both $T_{\rm bb}$ and $R_{\rm bb}$ as free parameters. 

\subsubsection{3FHL J1626.3-4915}
Part of the positional error ellipse of 3FHL J1626.3-4915 has been covered by a {\it Chandra} ACIS-I observation 
(Obs. ID. 13287) on 2012 June 16 for an effective exposure of $\sim10$~ks. 
It is also partially covered by an {\it XMM-Newton} observation (Obs ID. 0403280201) on 2007 February 14. However, this observation was seriously contaminated by 
high background. After removing these contaminated time intervals, an effective exposure of $\sim6$~ks is remained in the {\it XMM-Newton} data. 
In the {\it Chandra} observation, two sources, J16263\_X1 and J16263\_X2, are detected. 
The brighter one J16263\_X1 can also be detected in the short {\it XMM-Newton} exposure. The difference of its flux in these two frames is only $\sim2\sigma$. 
For J16263\_X2, it is below the detection threshold in the {\it XMM-Newton} observation. This places a limit of $>3.7\sigma$ on its long-term 
variability. Another source J16263\_X3, which is not covered by the FoV of the {\it Chandra} observation, is detected by {\it XMM-Newton}. 
From the sources serendipitously 
detected in these observations, the probability of chance coincidence is found to be $P\left(n\geq 1\right)\sim99\%$ and $P\left(n\geq 1\right)\sim61\%$ in the 
{\it Chandra} and {\it XMM-Newton} observation respectively. For computing the $F_{x}$ of 
all three detected sources as given in Table~4, we adopt a column absorption of $n_{H}=1.9\times10^{22}$~cm$^{-2}$ based on the HI estimate and 
their net count rates. 

$\sim87$ net counts are collected from J16263\_X1 altogether from {\it Chandra} and {\it XMM-Newton} data. This allows us to carry out a more detailed analysis. 
We found that its X-ray spectrum can be 
fitted equally well with both absorbed power-law ($\chi^{2}=16.31$ for 23 d.o.f.) and absorbed blackbody models ($\chi^{2}=15.78$ for 23 d.o.f.). 
The best-fit power-law model yields a column absorption of 
$n_{H}=6.5^{+10.3}_{-6.5}\times10^{21}$~cm$^{-2}$, a photon index of $\Gamma_{x}=2.0^{+0.7}_{-0.6}$ and an unabsorbed flux $F_{x}=2.7^{+3.8}_{-0.9}\times10^{-13}$ 
in 0.3-10~keV. For the best-fit blackbody model, it yields $n_{H}<3.4\times10^{21}$~cm$^{-2}$, a temperature $kT=0.9\pm0.1$~keV, emitting area with a radius 
of $R=4.5^{+2.1}_{-4.5}d_{kpc}$~km, and an unabsorbed flux $F_{x}=(1.2\pm0.2)\times10^{-13}$ in 0.3-10~keV. The goodness-of-fit for both models are 
comparable (see Table~6)

For J16263\_X3, there are $\sim130$ net counts collected by {\it XMM-Newton}. By fitting its spectrum with an absorbed power-law, we obtain 
the best-fit results of $n_{H}=8.4^{+2.4}_{-2.0}\times10^{21}$~cm$^{-2}$, $\Gamma_{x}=4.0^{+0.7}_{-0.6}$ and 
an unabsorbed flux $F_{x}\simeq1.9\times10^{-12}$ in 0.3-10~keV. On the other hand, the best-fit blackbody model yields 
$n_{H}=1.6^{+1.9}_{-1.5}\times10^{21}$~cm$^{-2}$, $kT=0.4\pm0.1$~keV, an emitting area with a radius
of $R=68^{+40}_{-21}d_{\rm kpc}$~m, and an unabsorbed flux $F_{x}\sim1.2\times10^{-13}$ in 0.3-10~keV. Although the power-law model 
yields a better goodness-of-fit ($\chi^{2}=10.51$ for 23 d.o.f.), its photon index is too steep to account for any reasonable non-thermal emission scenario. 

We have identified the optical/IR counterparts of J16263\_X1 and J16263\_X2. For J16263\_X1, after correcting the 
extinction with $A_{v}=8.9$ by assuming the total Galactic HI column density, its optical/IR counterpart can be described by a 
blackbody of $T_{\rm bb}\sim2000$~K and $R_{\rm bb}\sim1.1R_{\odot}d_{\rm kpc}$. However, its SED only has two data points and do not allow us to 
properly constrain its blackbody parameters and compute the uncertainties. For J16263\_X2, we found that the adopted $A_{v}=8.9$ might over-correct the 
extinction and results in an unphysical high blackbody temperature. On the other hand, a possible counterpart of J16263\_X2 is identified by {\it Gaia} and 
place an estimate on its distance to be 3.6~kpc.

\subsubsection{3FHL J1747.2-2822}
3FHL J1747.2-2822 is located along the line-of-sight towards the Galactic centre. 
The 95\% $\gamma-$ray positional error ellipse of 3FHL J1747.2-2822 has been covered by two {\it Chandra} observations: 
Obs.IDs.: 944 (2000 March 29; 100~ks), 11795 (2010 July 29; 98~ks) and five {\it XMM-Newton} observations 
Obs.IDs: 0802410101 (2018 April 2; 99~ks), 0694641401 (2012 September 30; 34~ks), 
0694641301 (2012 September 26; 47~ks), 0694640601 (2012 September 6; 41~ks), 
and 0203930101 (2004 September 4; 33~ks). 

Using the {\it Chandra} observation in 2010, eight X-ray sources were detected within
the $\gamma-$ray ellipse. In the 2000 {\it Chandra} observation, only J17472\_X1, J17472\_X2, J17472\_X4, J17472\_X5 and J17472\_X6 can be detected. 
In comparing these two frames, J17472\_X5 is found to be variable at the level of $4.3\sigma$. 
On the other hand, the non-detections of J17472\_X3, J17472\_X7 and J17472\_X8 place limits on their variabilities 
to be $>1.9\sigma$, $>1.1\sigma$ and $>0.3\sigma$ respectively. 
In all the {\it XMM-Newton} observations, only J17472\_X1 can be detected. Its flux as measured by these observations 
are all consistent with that obtained in {\it Chandra} observation. 

Their unabsorbed $F_{x}$ given in Table~4 are estimated by their count rate detected in the 2010 {\it Chandra} observation with 
the total Galactic HI column density $n_{H}=1.2\times10^{22}$~cm$^{-2}$ in that direction.
Since 3FHL J1747.2-2822 lies along the direction towards the Galactic centre,  
the spatial density of the X-ray sources is rather high. Using the 
serendipitous X-ray sources detected in these data, $P\left(n\geq 1\right)$ is estimated to be as high as 
$\sim100\%$ in all observations. 

Among the detected X-ray sources, J17472\_X1 is the brightest and is
detected by all observations. However, we found that it coincides with the giant molecular cloud Sgr B2. The association is further confirmed by the detection of 
the iron line at 6.4~keV which is likely to be originated from the interaction between the hard X-rays from the Sgr~A* and the cloud \citep{dogiel2015}. 
Therefore, J17472\_X1 is not the main interest for this work and will not be further concerned. 
For the other seven X-ray sources, their net counts are all less than 50 and no further analysis will be performed. 

We have also identified an optical/IR counterpart of J17472\_X8. Using an extinction of $A_{v}=5.7$ as inferred from the total Galactic HI 
column density, a blackbody fitting to the extinction-corrected SED yields a temperature of $T_{\rm bb}=1900\pm200$~K and an emitting area 
with a radius of $R_{\rm bb}=11\pm2d_{\rm kpc}$$R_{\odot}$. 

\subsubsection{3FHL J0737.5+6534}
3FHL J0737.5+6534 is located very far away from the Galactic plane with $b\sim29^{\circ}$. Its $\gamma-$ray positional error ellipse has been 
covered by three {\it Chandra} observations with Obs.IDs: 4630 (2004 December 22; 50~ks), 4629 (2004 October 3; 45~ks) and 
4628 (2004 August 23; 47~ks) and four {\it XMM-Newton} observations with Obs.IDs: 
0729560901 (2014 April 4; 2.7~ks), 030186301 (2005 October 11; 9.7~ks), 
0164560901 (2004 September 12; 56~ks) and 0150651101 (2003 April 30; 4.3~ks). 

Eight X-ray sources have been found within its $\gamma-$ray error ellipse by using the longest {\it Chandra} observation at 2004 December. 
Based on their count rates obtained from this observation and the adopted column absorption of 
$n_{H}=4.5\times10^{20}$~cm$^{-2}$ which is consistent with the total Galactic HI column density in that direction, we estimated their 
unabsorbed $F_{x}$ in Table~4. J07375\_X1 is the brightest X-ray sources among them. 
Its flux is found to be significantly variable at a level as high as $\sim12\sigma$. The shortest timescale of its 
X-ray variability found in this study is $\sim4$ months. 
J07375\_X2 and J07375\_X5 also exhibit long-term X-ray varibility at the level up to $8\sigma$ and $4\sigma$ among these observations. 

J07375\_X1 and J07375\_X2 have 1096 and 454 net counts collected from all these archival data respectively, and therefore we have carried 
out a more detailed analysis. Their photon statistics from each observation are high enough to allow us performing multi-epoch spectral analysis. 
By fixing $n_{H}$ at $4.5\times10^{20}$~cm$^{-2}$, the best-fit parameters in each epoch of J07375\_X1 and J07375\_X2 are summarized in Table~7 
and Table~8 respectively. Evidences for spectral variabilities are found from both sources. 

\begin{table}
	\begin{center}
	\begin{tabular}{c c c c c}
	\hline
	\hline
	Obs. Date             & $\chi^{2}$ & dof & $\Gamma_{x}$   & $F_{\rm 0.3-10 keV}^{\rm unabs}$             \\
	                    &                  &     &            & ($10^{-13}$ ergs cm$^{-2}$ s$^{-1}$) \\
		\hline
		2003-04-30 & 16.71            & 17  & $2.07_{-0.63}^{+1.21}$ & $0.32_{-0.07}^{+0.11}$   \\
		\hline
		2004-08-23 & 8.64            & 15  & $1.67\pm{0.11}$        & $0.65_{-0.05}^{+0.06}$   \\
		\hline
		2004-09-12 & 11.37            & 20  & $2.01\pm{0.08}$        & $0.43\pm{0.02}$          \\
		\hline
		2004-10-03 & 6.33            & 15  & $1.84\pm{0.10}$        & $0.79_{-0.05}^{+0.06}$   \\
		\hline
		2004-12-22 & 17.25            & 16  & $1.88\pm{0.11}$        & $0.58\pm{0.04}$          \\
		\hline
		2005-10-11 & 5.74            & 18  & $2.56_{-0.40}^{+0.43}$ & $0.21\pm{0.04}$          \\
		\hline
		2014-04-04 & 8.48            & 14  & $1.61_{-0.51}^{+0.53}$ & $0.39_{-0.13}^{+0.18}$   \\
		\hline

	\end{tabular}
	\end{center}
	\caption{X-ray spectral properties of J07375\_X1 at different epochs.}
\end{table}

\begin{table}
	\begin{center}
	\begin{tabular}{c c c c c}
	\hline
	\hline
	Obs. Date             & $\chi^{2}$ & dof & $\Gamma_{x}$   & $F_{\rm 0.3-10 keV}^{\rm unabs}$             \\
	                    &                  &     &            & ($10^{-13}$ ergs cm$^{-2}$ s$^{-1}$) \\
		\hline
		2004-08-23 & 10.08            & 15  & $1.00\pm{0.21}$        & $0.20\pm{0.04}$          \\
		\hline
		2004-10-03 & 13.30            & 13  & $1.39_{-0.23}^{+0.26}$ & $0.20_{-0.03}^{+0.04}$   \\
		\hline
		2004-12-22 & 13.29            & 14  & $0.98\pm{0.18}$        & $0.36_{-0.05}^{+0.06}$   \\
		\hline
		2003-04-30 & 9.97            & 18  & $1.73_{-0.66}^{+0.73}$ & $0.21_{-0.09}^{+0.12}$   \\
		\hline
		2004-09-12 & 29.06            & 20  & $1.19_{-0.18}^{+0.19}$ & $0.18_{-0.02}^{+0.03}$   \\
		\hline
		2005-10-11 & 11.10            & 13  & $0.04_{-0.83}^{+0.87}$ & $0.25_{-0.16}^{+0.20}$   \\
		\hline
		2014-04-04 & 1.43            &  4  & $1.37_{-1.37}^{+2.17}$ & $0.27_{-0.20}^{+21.34}$  \\
		\hline

	\end{tabular}
	\end{center}
	\caption{X-ray spectral properties of J07375\_X2 at different epochs.}
\end{table}


Among eight X-ray sources, six of them have optical/IR counterparts identified. The results of blackbody fitting to their SEDs are summarized in Table~5. 
Their inferred low temperatures and small radii suggest they can possibly be late-type stars. For J07375\_X3, counterpart has also been found by 
{\it Gaia} which suggests a distance of 1.6~kpc.
 

\section{Summary \& Discussions} 
With an optimal set of features selected by RFE algorithm (see Table~1 \& Figure~1),
a supervised classification model is built from a training set of labeled PSR/NON\_PSR 3FHL objects. 
Using this model, we have selected 27 PSR-like objects with a nominal accuracy of $\sim98\%$ 
from the unknown 3FHL sources for identification campaign (see Table~3).
Utilizing the archival X-ray data, we have found X-ray counterparts from 10 3FHL PSR-like candidates (see Table~4 \& Figure~4). These 
identifications allows us to systematically constrain the positions of the potential X-ray counterparts to arcsecond accuracies, estimate the X-ray to $\gamma-$ray 
flux ratios $F_{x}/F_{\gamma}$ and temporal variabilities. 
Except for J18007\_X8 and J17472\_X8, the $F_{x}/F_{\gamma}$ for all the other X-ray sources conform with that for 
the known pulsars detected in the energy range of 10~GeV to 2 TeV. For the sources with their X-rays found to be 
significantly varying in a given observation window and/or across different epochs, their flux variabilities make them less likely to be a typical pulsar 
which has rather stable X-ray emission. On the other hand, we cannot exclude the possibility of these variable X-ray sources as $\gamma-$ray binaries.
Also, their X-ray positions enable us to search for the optical/IR counterparts and estimate the surface temperatures and
sizes of the possible companion stars by assuming a blackbody model (see Table 6).

For those have more than 50 net counts collected from the archival X-ray data, we have carried out more detailed analysis. Among them, J18233\_X1 
which is associated with 3FHL~J1823.3-1339 is one of the most interesting source. They are very likely to be the X-ray and $\gamma-$ray counterparts of 
the globular cluster Mercer~5 (cf. Figure~6). The association between J18233\_X1 and Mercer 5 is supported by the consistency between the column absorption 
obtained from the X-ray spectral fitting and that deduced from the optical extinction. On the other hand, the association between 3FHL~J1823.3-1339 
and Mercer 5 is suggested by the agreement between its $\gamma-$ray luminosity $L_{\gamma}$ at the distance of the globular cluster and the general trend 
of $L_{\gamma}-$[Fe/H] as observed in the $\gamma$-ray globular cluster population \citep{hui2011}. 

Because of the frequent stellar encounters, globular clusters are efficient in producing compact binaries, including millisecond pulsars (MSPs), through 
dynamical interactions \citep{pooley2003, hui2010}. 
It is a general consensus that the $\gamma-$ray emission from a globular cluster is originated from its MSPs. Therefore, we speculate that Mercer 5 is 
hosting a MSP population awaited to be discovered. Pulsar searches targeted at this cluster are encouraged to examine this assertion. 

There are two different scenarios in explaining the $\gamma-$ray emission mechanism of a globular cluster. While their $\gamma$-rays 
can be collectively contributed by the magnetospheric radiation from the MSPs \citep{abdo2010}, it is possible that the inverse Compton 
scattering (ICS) between the relativistic pulsar wind and the ambient soft photons can result in the observed $\gamma$-rays \citep{cheng2010}.
The ICS scenario is suggested by the correlation between $L_{\gamma}$ and the energy densities of the ambient soft photon fields \citep{hui2011}. 
Such scattering can boost the soft photons to an energy $>10$~GeV \citep{cheng2010}. 
As the $\gamma$-ray spectrum of a pulsar typically has an exponential cut-off at a few GeV, their magnetospheric radiation is 
unlikely to have significant contribution in the hard $\gamma$-ray band. Therefore, the globular clusters detected at energies $>10$~GeV can help 
us to constrain the parameters of the ICS model \citep{cheng2010}. Although there are 30 $\gamma-$ray globular clusters have been identified 
in the {\it Fermi} LAT 8 years point source catalog \citep{fermi2019}, only two of them, 47~Tuc and Terzan~5, are included in the 3FHL catalog. 
On the other hand, Figure~A3 in \citet{deMenezes2019} shows that 2MS-GC01, NGC6440 and NGC2808 seem to have emission above 10 GeV. 
The survey with the upcoming CTA holds the potential in further expanding the population of hard $\gamma$-ray globular clusters. 

Apart from MSPs, globular clusters also host different classes of compact X-ray binaries (e.g. low-mass X-ray binaries, cataclysmic variables). 
Ascribing to the relatively poor spatial resolution of {\it XMM-Newton}, the X-ray counterpart of Mercer 5, J18233\_X1, identified in this work 
is likely resulted from a blend of unresolved X-ray point source population. Its X-ray spectrum can be well-described by a power-law model 
with $\Gamma_{x}\sim1.1$ (Figure~7) which is apparently harder than the faint unresolved X-ray populations found in the other clusters \citep{hui2009}. 
A spectral imaging analysis with high spatial resolution by {\it Chandra} is necessary to resolve and classify the X-ray binaries in Mercer~5. 

Another interesting identification in our campaign is 3FHL~J1405.1-6118 and its promising X-ray counterpart J14051\_X1. The X-ray spectrum of 
J14051\_X1 can be described by a power-law of $\Gamma_{x}\sim2.7$ with a large column absorption $n_{H}\sim1.5\times10^{23}$~cm$^{-2}$ (Figure~10). 
Such large X-ray absorptions are commonly seen in the high-mass X-ray binaries (HMXBs) \citep{paul2017}. Very recently, a $\gamma$-ray periodic modulation of 
$P_{b}\sim13.7$~days in 0.2-500 GeV has been discovered which makes 3FHL~J1405.1-6118 (= 4FGL J1405.1-6119) the third $\gamma-$ray binary found 
from the initial discovery of periodic modulation of the LAT light curve \citep{corbet2019}.  
X-ray modulation of J14051\_X1 at the same period has also been found by {\it Swift} XRT 
(cf. Figure~4 in \citet{corbet2019}). Taking the phase zero at MJD~56498.7, {\it Chandra} exposure used in our work corresponds the orbital phase of 
$\sim0.07-0.08$ which is not included in \citet{corbet2019}. Using our best-fit spectral model and with the aid of PIMMS, the flux observed by 
{\it Chandra} translates into a {\it Swift} XRT count rate of $\sim10^{-3}$~cts/s which is consistent with that in the phase interval of $\sim0.03-0.25$ 
as reported in the Table~1 of \citet{corbet2019}. 

For the optical/IR counterpart of 3FHL~J1405.1-6118, the blackbody fitting to its SED yields a temperature of $T\sim2000$~K and a radius of $R\sim7R_{\odot}$ 
in the case that we adopt $A_{v}=8.5$, $d=1$~kpc and with both $T_{\rm bb}$ and $R_{\rm bb}$ as free parameters. On the other hand, 
based on the near-IR spectroscopy, \citet{corbet2019} identify the counterpart as an O6 III star. To examine whether this inference can be consistent with 
our photometric result, we redo the blackbody fitting with $T_{\rm bb}$ fixed at 40000~K which is typical for O stars and adopt the mean derived extinction of 
$A_{v}=31.6$ ($E\left(B-V\right)=10.2$) 
reported by \citet{corbet2019}. At a distance of $d=7.7$~kpc (Corbet et al. 2019), it yields a radius of $R_{\rm bb}=12.9\pm0.8R_{\odot}$ 
which is consistent with the expected size for an O6 III star.

While \citet{corbet2019} have found the orbital period of $\sim13.7$~days, the X-ray light curve observed by {\it Chandra} suggests a 
periodicity candidate at $P\sim1.4$~hrs (Figure 11). For the HMXBs with the X-ray pulses from the neutron stars detected, their spin periods span a range from 
$\sim0.03$~s to $\sim4$~hrs \citep{liu2006}. Therefore, this signal can possibly be originated from the neutron star 
rotation. Deeper follow-up observations are strongly encouraged to examine this putative signal. 

Apart from the periodic signal candidate, this short {\it Chandra} observation also reveals a putative extended X-ray feature associated with J14051\_X1 
at a significance of $\sim4\sigma$ (Figures 12 \& 13). 
A deeper observation is required to confirm its spatial nature with higher signal-to-noise ratio and examine if there is any spectral variation across it. 
Evidences of such X-ray features have been found from a number of $\gamma$-ray binaries, including PSR B1259-63/LS 2883 \citep{pavlov2015}, 
LS I+61 303 \citep{paredes2007}, and LS 5039 \citep{durant2011}. Except for PSR B1259-63/LS 2883, the nature of the compact objects 
for the other $\gamma-$ray binaries remain unknown. For PSR B1259-63/LS 2883, its extended X-ray feature can be resulted from synchrotron radiation 
emitted by the relativistic particles accelerated at the shock between the pulsar wind and the massive star outflow \citep{tavani1997}.  
On the other hand, if the $\gamma$-ray binary is powered by a microquasar, the extended X-ray nebula can be originated from the relativistic particles produced 
by the Blandford-Znajek process \citep{bz1977} or from an MHD jet. Pulsar searches of 3FHL~J1405.1-6118 can help to discriminate these two 
competing scenarios. 
 
\section*{Acknowledgements}
CYH is supported by the National Research Foundation of Korea through grant 2016R1A5A1013277 and 2019R1F1A1062071.
JL is supported by National Research Foundation of Korea grant funded by the Korean Government (NRF-2019H1A2A1077350-Global Ph.D. Fellowship Program), 2016R1A5A1013277, 2019R1F1A1062071 and BK21 plus Chungnam National University.
KLL is supported by the Ministry of Science and Technology of Taiwan through grant 108-2112-M-007-025-MY3. 
SK is supported by BK21 plus Chungnam National University, National Research Foundation of Korea grant 2016R1A5A1013277 and 2019R1F1A1062071.
KO is supported by National Research Foundation of Korea grant funded by the Korean Government (NRF-2019H1A2A1077058-Global Ph.D. Fellowship Program), 2016R1A5A1013277, 2019R1F1A1062071 and BK21 plus Chungnam National University.
Alex P. L. and Shengda L. are funded by the Science and Technology Development Fund, Macau SAR (No. 0019/2018/ASC).
AKHK is supported by the Ministry of Science and Technology of the Republic of China (Taiwan) through grants 105-2119-M-007-028-MY3 and
106-2628-M-007-005. JT is supported by the NSFC grants of the Chinese Government under 11573010, 11661161010, U1631103 and U1838102. 
KSC is supported by GRF grant under 17302315.












\bsp	
\label{lastpage}
\end{document}